\documentclass[sigconf]{aamas} 

\usepackage{balance} 

\usepackage{algorithm}
\usepackage{algorithmic}
\usepackage{amsmath}

\newtheorem{proposition}{Proposition}
\usepackage{verbatim}
\usepackage{cuted}
\usepackage{mathtools}
\usepackage{amsfonts,latexsym,color}
\usepackage{multirow}

\setcopyright{ifaamas}
\acmConference[AAMAS '23]{Proc.\@ of the 22nd International Conference
on Autonomous Agents and Multiagent Systems (AAMAS 2023)}{May 29 -- June 2, 2023}
{London, United Kingdom}{A.~Ricci, W.~Yeoh, N.~Agmon, B.~An (eds.)}
\copyrightyear{2023}
\acmYear{2023}
\acmDOI{}
\acmPrice{}
\acmISBN{}

\acmSubmissionID{434}

\title[AAMAS-2023 Formatting Instructions]{A Variational Approach to  Mutual Information-Based Coordination for Multi-Agent Reinforcement Learning}

\author{Woojun Kim}
\affiliation{
  \institution{KAIST}
  \city{Daejeon}
  \country{Korea}}
\email{woojun.kim@kaist.ac.kr}

\author{Whiyoung Jung}
\affiliation{
  \institution{KAIST}
  \city{Daejeon}
  \country{Korea}}
\email{wy.jung@kaist.ac.kr}

\author{Myungsik Cho}
\affiliation{
  \institution{KAIST}
  \city{Daejeon}
  \country{Korea}}
\email{ms.cho@kaist.ac.kr}

\author{Youngchul Sung}
\affiliation{
  \institution{KAIST}
  \city{Daejeon}
  \country{Korea}}
\email{ycsung@kaist.ac.kr}

\begin{abstract}
In this paper, we propose a new mutual information (MMI) framework for multi-agent reinforcement learning (MARL) to enable multiple agents to learn coordinated behaviors by regularizing the accumulated return with the simultaneous mutual information between multi-agent actions. By introducing a latent variable to induce nonzero mutual information between multi-agent actions and applying a variational bound, we derive a tractable lower bound on the considered MMI-regularized  objective function. The derived tractable objective can be  interpreted as  maximum entropy reinforcement learning combined with uncertainty reduction of other agents’ actions. Applying policy iteration to maximize the derived lower bound, we propose a practical algorithm named variational maximum mutual information multi-agent actor-critic (VM3-AC), which follows centralized learning with decentralized execution (CTDE). We evaluated VM3-AC for several games requiring coordination, and numerical results show that VM3-AC outperforms  other MARL algorithms in multi-agent tasks requiring high-quality coordination.
\end{abstract}

\keywords{Multi-Agent Reinforcement Learning; Coordination; Mutual Information}

\newcommand{\BibTeX}{\rm B\kern-.05em{\sc i\kern-.025em b}\kern-.08em\TeX}

\begin{document}


\pagestyle{fancy}
\fancyhead{}


\maketitle

\section{Introduction}
\label{sec:Introduction}

With the success of RL in the single-agent domain  \cite{2015Mnih,2015Lillicrap}, MARL is being  actively studied  and  applied to real-world problems such as  traffic control systems and connected self-driving cars, which can be  modeled as multi-agent systems requiring coordinated control \cite{2019Li, 2019Andriotis}.
The simplest approach to MARL is independent learning, which trains each agent independently while treating other agents as a part of the environment, but this approach suffers from the problem of non-stationarity of the environment.
A common solution to this  problem is to use  fully-centralized critic in the framework of centralized training with decentralized execution (CTDE) \cite{2019Oroojlooyjadid, 2018rashid,2017Lowe,2018Iqbal, jeon2022maser}. For example, MADDPG \cite{2017Lowe} uses a centralized critic to train a decentralized policy for each agent, and COMA \cite{2018Foerster} uses a common centralized critic to train all decentralized policies.
However, these approaches assume that decentralized policies are independent and hence  the joint policy is the product of each agent's policy.
Such  non-correlated factorization of the joint policy limits the agents to learn coordinated behavior due to negligence of the influence of other agents \cite{2019Wen,2019de}.
Recently, {\em mutual information} (MI)  between multiple agents' actions has been considered as an effective  intrinsic reward to promote coordination in MARL  \cite{2019Jaques}.  In \cite{2019Jaques}, MI between agents' actions is captured as social influence and the goal is to maximize the sum of accumulated return and social influence between agents' actions. It is shown that the social influence approach is effective for sequential social dilemma games. 
In this framework, however, causality between actions under coordination is required,  and it is not straightforward to coordinate multi-agents' simultaneous actions. In certain multi-agent games,  coordination of simultaneous actions of multiple agents is required to achieve cooperation for a common goal. For example, suppose that a pack of wolves tries to catch a prey. To catch the prey, coordinating simultaneous actions among the wolves is more effective than coordinating one wolf's action and other wolves' actions at the next time because the latter case causes delay in coordination.  
In this paper, we propose a new approach to the  MI-based coordination for MARL to coordinate simultaneous actions among multiple agents under the assumption of the knowledge of timing information among agents. Our approach is based on introducing a common latent variable to induce MI among simultaneous actions of multiple agents and on a variational lower bound on MI that enables tractable optimization.  Under the proposed formulation, applying policy iteration by redefining value functions, we propose the VM3-AC algorithm for MARL to learn coordination of simultaneous actions among multiple agents. Numerical results show its superior performance on cooperative multi-agent tasks requiring coordination.

\section{Related Work}
\label{sec:RelatedWork}

MI is a measure of dependence between two variables \cite{Cover:book} and  has been considered as an effective intrinsic reward for MARL \cite{2019Wang,2019Jaques}.  \cite{2015mohamed} proposed an intrinsic reward for empowerment by maximizing MI between agent's action and its future state.  \cite{2019Wang} proposed two intrinsic rewards capturing the influence based on a decision-theoretic measure and MI between an agent's current actions/states and other agents' next states. 
In particular, \cite{2019Jaques} proposed a social influence intrinsic reward, which basically captures  the mutual information between multiple agents' actions to achieve coordination, and showed that the social influence formulation yields good performance in sequential social dilemma environments. The difference of our approach from the social influence to MI-based coordination will be explained in Section \ref{sec:Background} and Section \ref{subsec:formulation}.

Some previous works approached correlated policies from different perspectives.  \cite{2020Liu} proposed  explicit modeling of correlated policies for multi-agent imitation learning, and \cite{2019Wen} proposed a recursive reasoning framework for MARL to  maximize the expected return by decomposing the joint policy into own policy and opponents' policies. Going beyond adopting correlated policies, our approach maximizes the MI between multiple agents' actions which is a measure of correlation.

In our approach, the MI between agents' action distributions is decomposed as the sum of each agent's action entropy and a variational term related to prediction of other agents' actions.  Hence, our framework can  be interpreted as enhancing correlated exploration by increasing the entropy of own policy \cite{2018Haarnoja} while decreasing the uncertainty about other agents' actions.  Some previous works proposed other techniques to enhance  correlated exploration \cite{2018zheng, 2019Mahajan2019}.  MAVEN addressed the poor exploration problem of QMIX by maximizing the mutual information between the latent variable and the observed trajectories \cite{2019Mahajan2019}. However, MAVEN does not consider the correlation among policies.

\section{Background}
\label{sec:Background}

\textbf{Setup}   ~We consider a Markov Game \cite{1994Littman}, which is an extention of Markov Decision Process (MDP) to multi-agent setting. An $N$-agent Markov game is defined by an environment state space $\mathcal{S}$, action spaces for $N$ agents $\mathcal{A}_1,\cdots,\mathcal{A}_N$, a state transition probability $p_{\mathcal{T}} : \mathcal{S} \times \boldsymbol{\mathcal{A}} \times \mathcal{S} \rightarrow [0,1]$, where $\boldsymbol{\mathcal{A}}=\prod_{i=1}^{N}\mathcal{A}_i$ is the joint action space, and a reward function $\mathcal{R}: \mathcal{S}\times \boldsymbol{\mathcal{A}} \rightarrow {\mathbb{R}}$. At each time step $t$, Agent $i$ with policy $\pi^i$ executes action $a_t^i\in \mathcal{A}_i$ based on  state $s_t\in \mathcal{S}$. The actions of all agents $\boldsymbol{a}_t=(a_t^1,\cdots,a_t^N)$ yield the  next state $s_{t+1}$ according to $p_{\mathcal{T}}$ and shared common reward $r_t$ according to $\mathcal{R}$ under the assumption of fully-cooperative MARL.
The discounted return is defined as $R_t = \sum_{n=t}^{\infty} \gamma^{n} r_{n}$, where $\gamma \in [0,1)$ is the discounting factor.

We assume CTDE incorporating the resource asymmetry between training and execution phases, widely considered in MARL \cite{2017Lowe,2018Iqbal,2018Foerster}.
Under CTDE, each agent can access all information including the environment state, observations and actions of other agents  in the training phase, whereas the policy of each agent is conditioned only on its own  observation $o_t^i$ in the execution phase.
The goal of fully cooperative MARL is to find the optimal joint policy $\boldsymbol{\pi}^*$ that maximizes the objective $J(\boldsymbol{\pi})=E_{\tau_0\sim\boldsymbol{\pi}} \big[R_0\big]$, where $\tau_t=(s_t,\boldsymbol{a}_t,s_{t+1},\boldsymbol{a}_{t+1},\cdots)$ and $\boldsymbol{\pi}=(\pi^1,\cdots,\pi^N)$ denotes the joint policy of all agents.

\begin{figure}[t]
\begin{center}
\begin{tabular}{ccc}
     \includegraphics[width=0.12\textwidth]{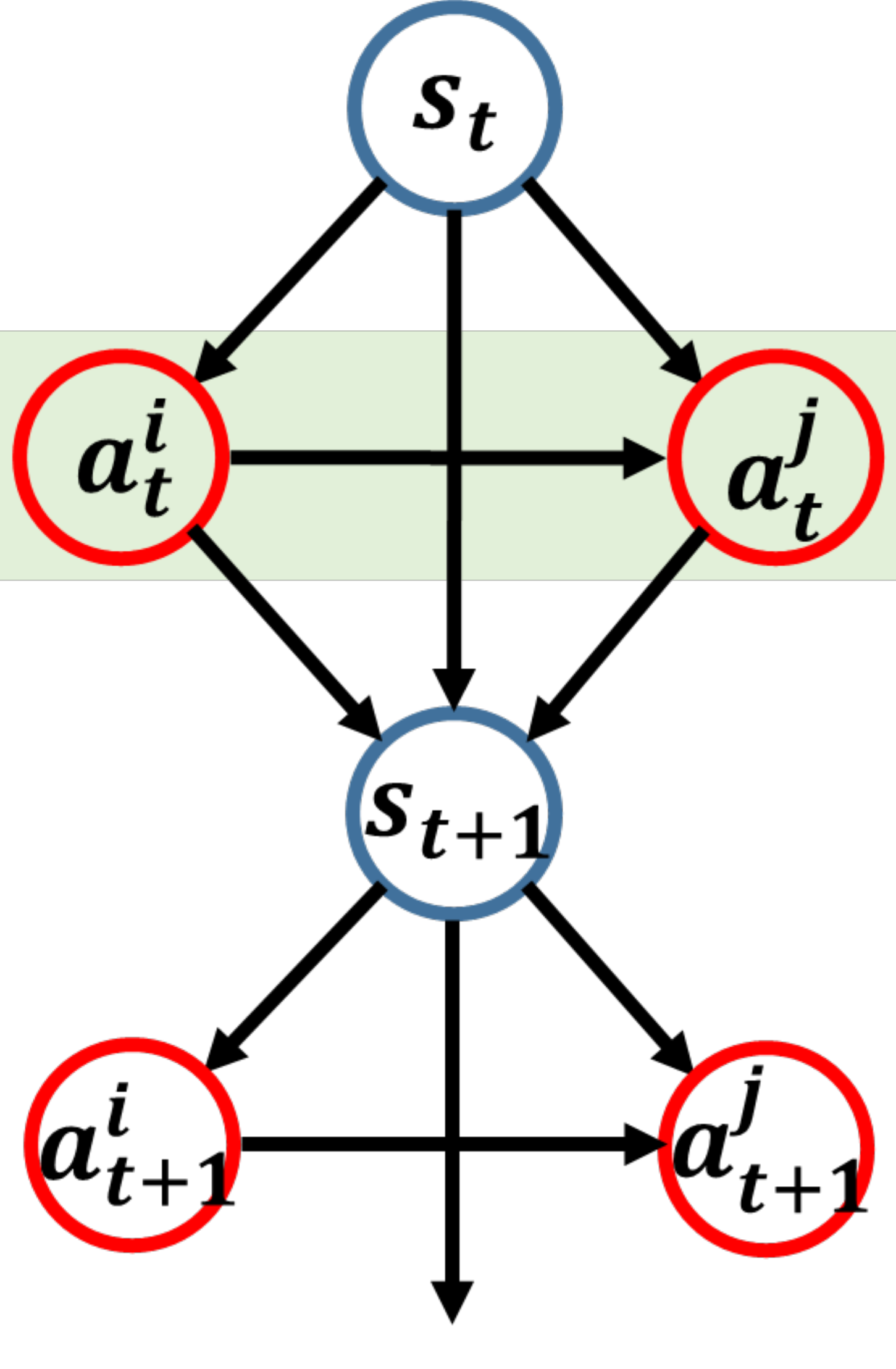} &
     \includegraphics[width=0.13\textwidth]{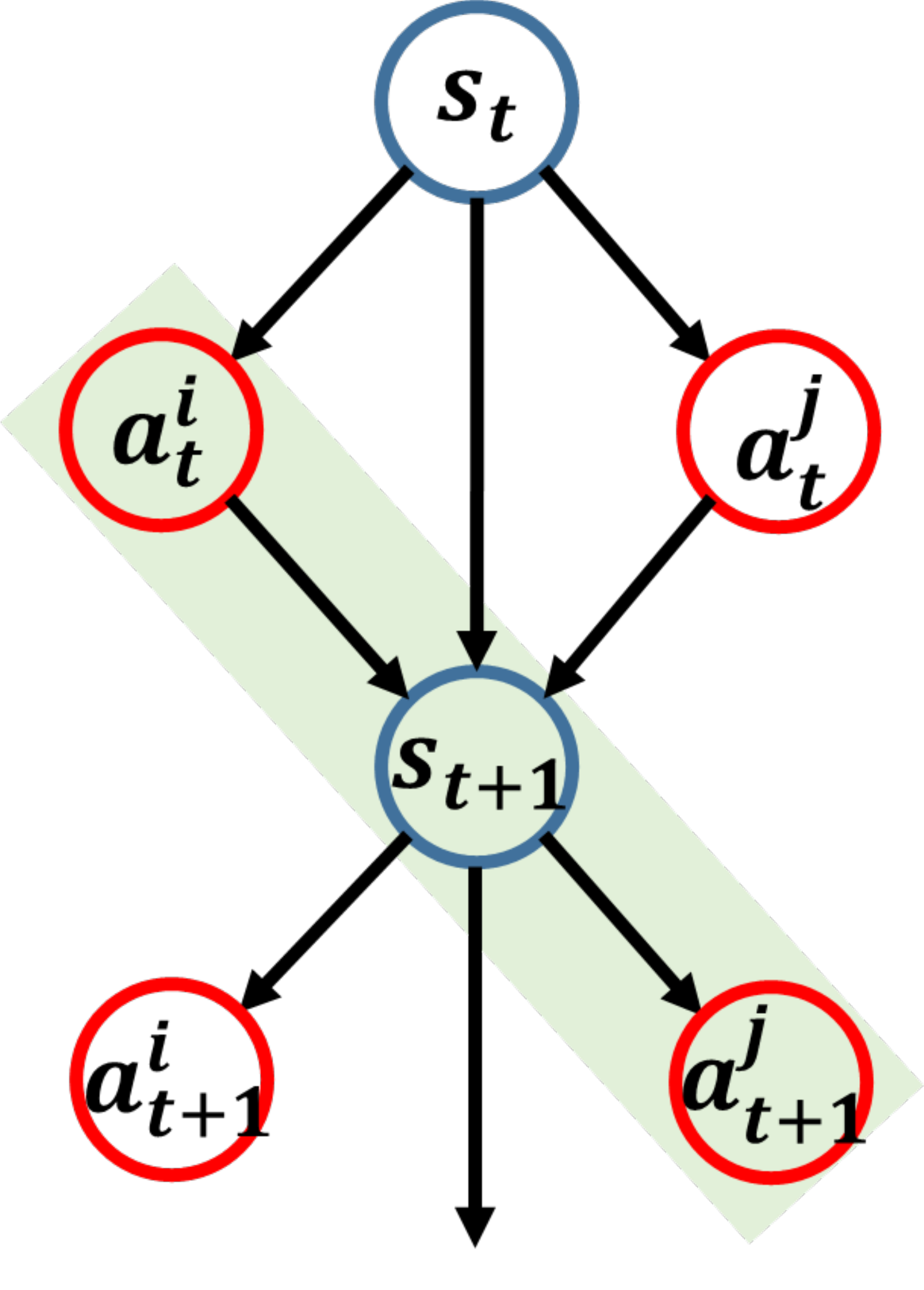} &
     \includegraphics[width=0.12\textwidth]{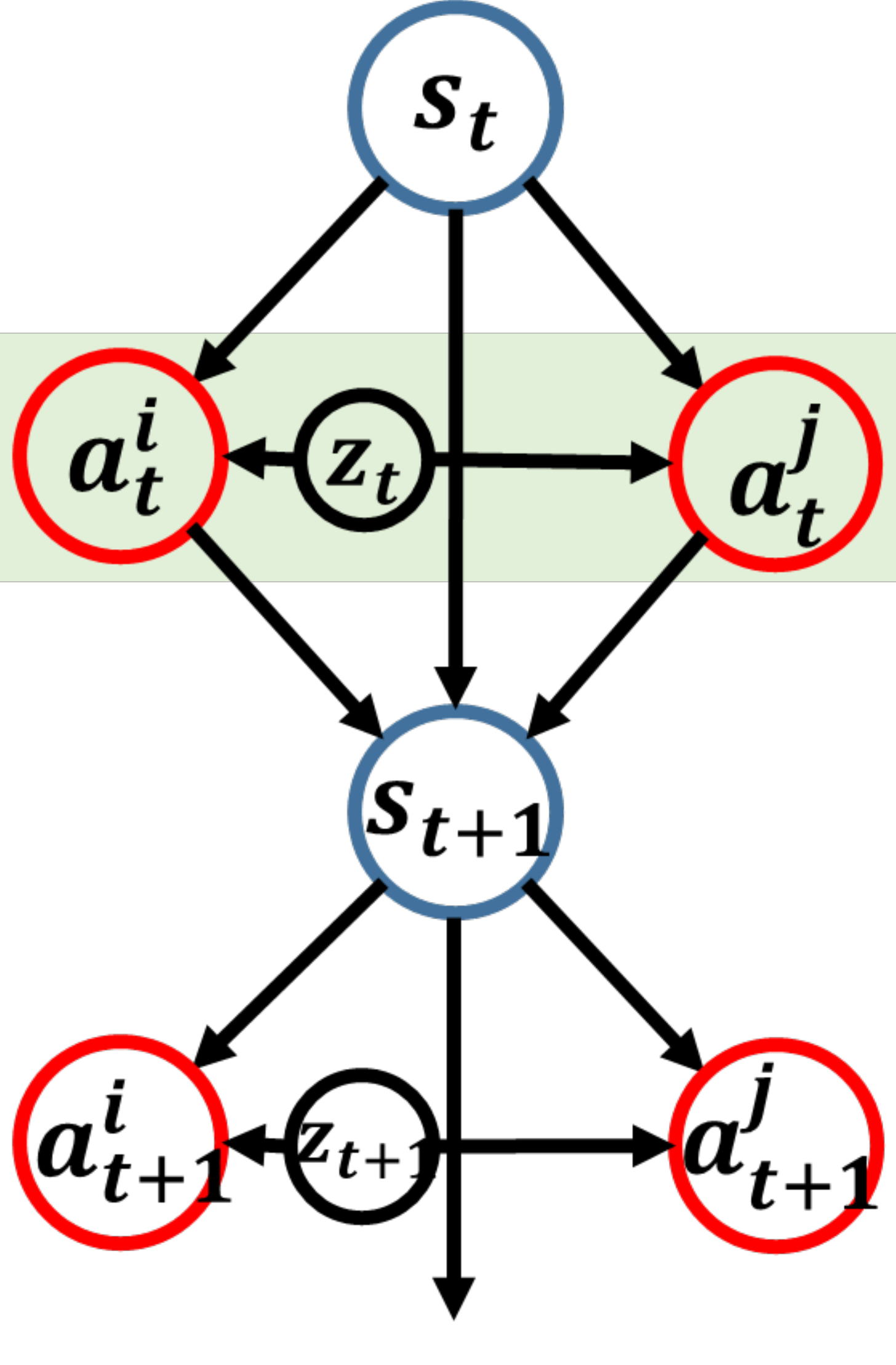} \\
     (a) & (b) & (c)
\end{tabular}
 \caption{Causal diagram: (a) basic social influence, (b) social influence of modeling other agents, and (c) the proposed approach}
\label{fig:CausalDiagram}
\end{center}
\end{figure}

\textbf{Mutual Information-Based Coordination for MARL} ~ MI  between agents' actions has been considered as an intrinsic reward to promote coordination in MARL \cite{2019Jaques}. Under this framework, one basically  aims to find the policy that maximizes the weighted sum of the  return and the MI between multi-agent actions.  Thus, the MI-regularized objective function for joint policy $\boldsymbol{\pi}$ is given by 
\begin{equation}\label{eq:obj_mmi}
        J(\boldsymbol{\pi})=\mathbb{E}_{\tau_0\sim\boldsymbol{\pi}} \Bigg[\sum_{t=0}^{\infty} \gamma^t\Big( r_t+\alpha \sum_{(i,j)|i\ne j}I(a_t^i;a_t^j|s_t)\Big) \Bigg],
\end{equation}
where $I(a_t^i;a_t^j|s_t)$ is the MI between  $a_t^i \sim \pi^i(\cdot|s_t)$ and $a_t^j \sim \pi^j(\cdot|s_t)$,  and $\alpha$ is the temperature parameter that controls the relative importance of the MI against the reward. It is known that by regularization with MI in the  objective function (\ref{eq:obj_mmi}), the policy of each agent is encouraged to coordinate with other agents' policies.   There are several approaches to implement  \eqref{eq:obj_mmi}. Under the social influence framework in \cite{2019Jaques}, the MI  is decomposed as 
\begin{align}
&I(a_t^i;a_t^j|s_t) 
= \int_{a_t^i,a_t^j} p(a_t^i,a_t^j|s_t)\log\frac{p(a_t^i,a_t^j|s_t)}{p(a_t^i|s_t)p(a_t^j|s_t)}\label{eq:MIfirstExp}\\
&= \int_{a_t^i} p(a_t^i|s_t)\int_{a_t^j}p(a_t^j|a_t^i,s_t)\log \frac{p(a_t^j|a_t^i,s_t)}{p(a_t^j|s_t)} \\ 
&= \int_{a_t^i} p(a_t^i|s_t) \underbrace{D_{KL}(p(a_t^j|a_t^i,s_t) || p(a_t^j|s_t))}_{\stackrel{\triangle}{=}~\text{social influence of agent $i$ on agent $j$}}, \label{eq:SIdecomp}
\end{align}
where $D_{KL}(\cdot||\cdot)$ is the Kullback-Leibler divergence. In this decomposition, influencing Agent $i$'s policy is given by $\pi^i=p(a_t^i|s_t)$ and influenced Agent $j$'s policy is given by $\pi^j=p(a_t^j|a_t^i,s_t)$.  The social influence is defined as the difference between $p(a_t^j|a_t^i,s_t)$ and $p(a_t^j|s_t)$. Hence, at time step $t$, influencing Agent $i$ acts first and then influenced Agent $j$ acts based on $a_t^i$ after Agent $i$ acts, as shown in Fig. \ref{fig:CausalDiagram}(a). This sequential dependence between actions prevents multiple agents from performing simultaneous actions, which is an assumption of most decentralized execution. In addition, the social influence approach needs a strategy for action ordering  because it divides all agents into a set of influencers and a set of influencees. One way to remove this action ordering is to model other agents \cite{2019Jaques}. In this case,  the causal influence of action $a_t^i$  of Agent $i$ at time $t$ on action $a_{t+1}^j$ of Agent $j$ at time $t+1$ is considered, as shown in Fig.  \ref{fig:CausalDiagram}(b), i.e., the social influence $D_{KL}(p(a_{t+1}^j|a_t^i,s_t) || p(a_{t+1}^j|s_t))$ instead of the influence term in \eqref{eq:SIdecomp} is considered based on modeling $p(a_{t+1}^j|a_t^i,s_t)$ so that actions $a_t^{i}$ and $a_t^j$ can be performed simultaneously without ordering. In this case, however, the actually considered MI is $I(a_t^i;a_{t+1}^j|s_t)$ and  is not the MI between $a_t^i$ and $a_t^j$ occurring at the same time.
In this paper, we propose a different approach to MI regularization which enables  simultaneous coordination between actions $a_t^i$ and $a_t^j$ both at time $t$ without action ordering.

\section{The Proposed Approach}
\label{Method}

We assume that the environment is fully observable, i.e., each agent can observe the environment state $s_t$ for theoretical development in this section, and will consider  partially observable environment for practical algorithm construction under CTDE in the next section.

\subsection{Formulation}\label{subsec:formulation}

Without explicit dependency between actions,  $\pi^i(a_t^i|s_t)$ and $\pi^j(a_t^j|s_t)$ are conditionally independent for given environment state $s_t$ and consequently the mutual information is always zero, i.e., $I(\pi^i(\cdot|s_t);\\ \pi^j(\cdot|s_t))=0$. Then, the MI-regularized objective function \eqref{eq:obj_mmi} reduces to the standard MARL objective of only the accumulated return. In order to circumvent this difficulty, we propose a novel method to induce MI between actions. Our approach for inducing MI between concurrent two actions $a_t^i$ and $a_j^t$ of Agents $i$ and $j$ at time $t$ is to introduce a latent  variable $Z_t$, as shown in Fig. \ref{fig:CausalDiagram}(c). 
We assume that the latent variable $Z_t$ has a prior distribution $p_Z(z_t)$ and  that  actions $a_t^i$ and $a_t^j$ are generated from the  state variable $s_t$ and the latent random variable $Z_t$. Thus, Agent $i$'s action $a_t^i$ at time $t$ is drawn from the policy distribution of Agent $i$ as
\begin{equation}\label{eq:atiFormula}
    a_t^i \sim \pi^i(\hspace{0.1em}\cdot \hspace{0.1em}|S_t=s_t,Z_t),~~~i=1,2,\cdots,N,
\end{equation}
where we use the upper case for random variables and the lower case for their realizations  in the conditioning input terms for notational clarification. Then, even in  case of deterministic policy, 
there is randomness in $a_t^i$ for given $S_t=s_t$ due to the random input $Z_t$ since  a function of random variable is a random variable.  In case of stochastic policy, there is additional  randomness in $a_t^i$ for given $S_t=s_t$ due to stochasticity of the policy itself.  One can view the randomness due to $Z_t$ as a perturbation to nominal $a_t^i$ for given $S_t=s_t$.  
With the common perturbation-inducing variable $Z_t$ to all agents' policies,  two random variables $a_i^t$ and $a_j^t$ conditioned on $S_t=s_t$ are correlated due to  common $Z_t$, and then nonzero MI   $I(a_t^i;a_t^j|s_t)$ between concurrent $a_t^i$ and $a_t^j$ is  induced. We aim to exploit this correlation for action coordination and correlated exploration in the training phase. (See Appendix A for a simple example and explanation of our basic idea with the simple example.)

With nontrivial MI  $I(a_t^i;a_t^j|s_t)$, we now express this MI. First, note  in \eqref{eq:SIdecomp}
that we need $p(a_t^j|a_t^i,s_t)$ to compute the MI but we do not want to use  $p(a_t^j|a_t^i,s_t)$ directly because $p(a_t^j|a_t^i,s_t)$ requires Agent $j$ to know the action $a_t^i$ of Agent $i$. For this, we adopt a variational distribution $q(a_t^j|a_t^i,s_t)$ to estimate $p(a_t^j|a_t^i,s_t)$ and derive a lower bound on the MI  $I(a_t^i;a_t^j|s_t)$ as follows:
\begin{align}
& I(a_t^i;a_t^j|s_t)= \int_{a_t^i,a_t^j} p(a_t^i,a_t^j|s_t)\log\frac{p(a_t^i,a_t^j|s_t)}{p(a_t^i|s_t)p(a_t^j|s_t)} \nonumber \\
&=  \int_{a_t^i,a_t^j} p(a_t^i,a_t^j|s_t)\log\frac{p(a_t^i|s_t)p(a_t^j|a_t^i,s_t)q(a_t^j|a_t^i,s_t)}{p(a_t^i|s_t)p(a_t^j|s_t)q(a_t^j|a_t^i,s_t)}\nonumber\\
    &= \mathbb{E}_{p(a_t^i,a_t^j|s_t)}\left[ \log \frac{q(a_t^j|a_t^i,s_t)}{p(a_t^j|s_t)}\right] \nonumber \\ &~~~~~~~~ \times \mathbb{E}_{p(a_t^i|s_t)}\left[D_{KL}(p(a_t^j|a_t^i,s_t)\|q(a_t^j|a^i,s_t)) \right] \nonumber \\
    &\geq H(a_t^j|s_t)+ \mathbb{E}_{p(a_t^i|s_t)p(a_t^j|a_t^i,s_t)}\left[\log q(a_t^j|a_t^i,s_t) \right],  \label{eq:MILB1}
\end{align}
where the last inequality in \eqref{eq:MILB1} holds because the KL divergence is always non-negative. Note that $H(a_t^j|s_t)$ is the entropy of  $a_t^j$ given $s_t$, i.e., the entropy of the following marginal distribution of $a_t^j$ in our case:
\begin{equation}  \label{eq:piTilde}
    \tilde{\pi}^j(a_t^j|s_t) := \int_{z_t} \pi^j(a_t^j|S_t=s_t,Z=z_t)p_Z(z_t)dz_t.
\end{equation}
  For the variational distribution  $q(a_t^j|a_t^i,s_t)$ we consider a class of distributions $\mathcal{Q}$, i.e., $ q(a_t^j|a_t^i,s_t) \in \mathcal{Q}$.
The lower bound \eqref{eq:MILB1} becomes tight when $q(a_t^j|a_t^i,s_t)$  approximates $p(a_t^j|a_t^i,s_t)$ well, i.e., $D_{KL}(p(a_t^j|a_t^i,s_t)\|q(a_t^j|a^i,s_t))$ is small.  Note that in our expansion, the lower bound on the MI  $I(a_t^i;a_t^j|s_t)$ is expressed as {\em the sum of the action entropy $H(a_t^j|s_t)$ and the negative of the cross entropy of $q(a_t^j|a_t^i,s_t)$ relative to $p(a_t^j|a_t^i,s_t)$} averaged over $p(a_t^i|s_t)$.  Using the symmetry of MI, we can rewrite the lower bound as
\begin{align}
    &I(a_t^i;a_t^j|s_t) \geq \frac{1}{2} \Big\{ H(a_t^i|s_t)+ H(a_t^j|s_t) \nonumber \\ 
    &+ \mathbb{E}_{p(a_t^i,a_t^j|s_t)}\left[\log q(a_t^j|a_t^i,s_t) + \log q(a_t^i|a_t^j,s_t)   \right] \Big\}. \label{eq:lb_mi2}
\end{align}
Then, our goal is to  maximize this lower bound of MI  by using a tractable approximation $q(a_t^i|a_t^j,s_t) \in \mathcal{Q}$. Our decompsition of MI based on the action entropy and the cross entropy is effective in our variational formulation for MI-based MARL.  Consider one of the cross entropy terms in the right-hand side (RHS) of  \eqref{eq:lb_mi2}:  $\mathbb{E}_{p(a_t^i,a_t^j|s_t)}[\log q(a_t^j|a_t^i,s_t)]$, which can be rewritten as
\begin{align}
\mathbb{E}_{p(a_t^i,a_t^j|s_t)}&[\log q(a_t^j|a_t^i,s_t)] =-\mathbb{E}_{p(a_t^i|s_t)} \biggl[ H(p(a_t^j|a_t^i,s_t) )
 \nonumber \\ & ~~~~~~~~~~~~~~~~~~~ +D_{KL}(p(a_t^j|a_t^i,s_t)||q(a_t^j|a_t^i,s_t))  \biggr],  \label{eq:crossEntropyTerm}
\end{align}
based on the well-known decomposition of the cross entropy. 
Hence, by maximizing  this cross entropy term, due to the negation  in (\ref{eq:crossEntropyTerm}) 
we can learn $\pi^i$ (generating $a_t^i$) and $\pi^j$ (generating $a_t^j$) so that  
the conditional entropy $H(p(a_t^j|a_t^i,s_t))$ of $a_t^j$ given $a_t^i$ is minimized, i.e., the two actions are more correlated to each other, and learn $q$ that  closely approximates the true $p(a_t^j|a_t^i,s_t)$, i.e., the $D_{KL}$ term in \eqref{eq:crossEntropyTerm} is minimized.

\subsection{Modified Policy Iteration}
\label{subsec:ModifiedPI}

Our algorithm construction is based on policy iteration.
In order to develop policy iteration for the proposed MI framework, we first replace the original MI-regularized objective function (\ref{eq:obj_mmi}) with the following tractable objective function
based on the variational lower bound (\ref{eq:lb_mi2}):
\begin{align}\label{eq:obj_mmi_bound}
 \hat{J}(\boldsymbol{\pi}, q)
        =&\mathbb{E}_{\scriptsize\begin{array}{c}
       \tau_0 \sim \boldsymbol{\pi}\\
        z_t\sim p_Z
        \end{array}}\Bigg[\sum_{t=0}^{\infty}  \gamma^t \Big(  r_t(s_t,\boldsymbol{a_t})  \nonumber \\ & +  \alpha N \sum_{i=1}^{N}H(a_t^i|s_t) + \alpha \sum_{i=1}^{N}\sum_{j\neq i}\log q(a_t^j|a_t^i,s_t) \Big) \Bigg],
\end{align}
where $\boldsymbol{\pi}=[\pi^1,\cdots,\pi^N]$ and $\pi^i$ is given by \eqref{eq:atiFormula} and $\boldsymbol{a}_t=[a_t^1,\cdots,a_t^N]$.  Then, we determine the individual objective function $\hat{J}^i(\pi^i,q)$  for Agent $i$ as the sum of the terms in (\ref{eq:obj_mmi_bound}) associated with Agent $i$'s policy $\pi^i$ or action $a_t^i$, given by {\small 
\begin{align}
    &\hat{J}^i(\pi^i,q) =  
\mathbb{E}_{\scriptsize\begin{array}{c}
       \tau_0 \sim \boldsymbol{\pi}\\
        z_t\sim p_Z
        \end{array}}
\Bigg[\sum_{t=0}^{\infty}\gamma^t \Big( \underbrace{ r_t(s_t,\boldsymbol{a_t}) + \beta  \cdot H(a^i|s_t)}_{(a)} \nonumber \\ 
& ~~~~ + \frac{\beta}{N}\sum_{j\neq i}\Big[\underbrace{\log q(a_t^i|a_t^j,s_t)+\log q(a_t^j|a_t^i,s_t)}_{(b)}\Big] \Big)\Bigg], \label{eq:obj_mmi_bound_agent}
\end{align}
}where $\beta = \alpha N$ is the temperature parameter.
Note that maximizing the term (a) in (\ref{eq:obj_mmi_bound_agent}) implies that each agent maximizes the weighted sum of the return and the action entropy, which can be interpreted as an extension of maximum entropy RL \cite{2018Haarnoja}  to multi-agent setting.
On the other hand, maximizing the term (b) with respect to $\pi^i$ and $q$ means that
we update the policy $\pi^i$  
so that the conditional entropy of $a_t^j$ given $a_t^i$ and the conditional entropy of $a_t^i$ given $a_t^j$ are reduced, as already mentioned below 
\eqref{eq:crossEntropyTerm}.
Thus, the objective function (\ref{eq:obj_mmi_bound_agent}) can be interpreted  as the {\em maximum entropy MARL objective combined with
action correlation or coordination}.  Hence, the proposed objective function (\ref{eq:obj_mmi_bound_agent}) can be considered as one implementation of the concept of {\em correlated exploration} in MARL \cite{2019Mahajan2019}.

Now, in order to learn policy $\pi^i$ to maximize the objective function (\ref{eq:obj_mmi_bound_agent}), we modify the policy iteration in standard RL. For this, we redefine the  value functions for Agent $i$  as
{\small\begin{align}
    &Q_i^{\boldsymbol{\pi}}(s,a) \triangleq \mathbb{E}_{\scriptsize\begin{array}{c}
       \tau_0 \sim \boldsymbol{\pi}\\
        z_t\sim p_Z
        \end{array}}\Bigg[ r_0 + \gamma V_i^{\boldsymbol{\pi}}(s_{1}) \Bigg| s_0=s, \boldsymbol{a}_0=\boldsymbol{a} \Bigg], \label{eq:value_function11}\\
    &V_i^{\boldsymbol{\pi}}(s)
    \triangleq \mathbb{E}_{\scriptsize\begin{array}{c}
       \tau_0 \sim \boldsymbol{\pi}\\
        z_t\sim p_Z
        \end{array}}\Bigg[ \sum_{t=0}^{\infty}\gamma^t \Big(r_{t} +\beta H(a_t^i| s_{t}) \nonumber \\ & \hspace{4ex} +\frac{\beta}{N}\sum_{j\neq i}\log q^{(i,j)}(a_{t}^i,a_{t}^j|s_{t})\Big) \Bigg| s_0=s \Bigg], \label{eq:value_function12} 
\end{align}}where $q^{(i,j)}(a_t^i,a_t^j|s_t)\triangleq q(a_t^i|a_t^j,s_t)q(a_t^j|a_t^i,s_t)$.
Then, the Bellman operator corresponding to  $V_i^{\boldsymbol{\pi}}$ and $Q_i^{\boldsymbol{\pi}}$ on the value function estimates $V_i(s)$ and $Q_i(s,\boldsymbol{a})$ is given by
\begin{align}\label{eq:bellmanMain}
    \mathcal{T}^{\boldsymbol{\pi}}Q_i(s,\boldsymbol{a}) &\triangleq r(s,\boldsymbol{a}) + \gamma \mathbb{E}_{s'\sim p}[V_i(s')], ~~~~ \mbox{where}
\end{align}
$V_i(s)= \mathbb{E}\Bigg[Q_i(s,\boldsymbol{a})
    -\beta \log \tilde{\pi}^i(a^i|s)   + \frac{\beta}{N} \sum_{j\neq i}\log q^{(i,j)}(a^i,a^j|s) \Bigg],$
and $\tilde{\pi}^i$ is the marginal distribution given in   \eqref{eq:piTilde}.
In the policy evaluation step, we compute the value functions  (\ref{eq:value_function11}) and (\ref{eq:value_function12}) by applying the modified Bellman operator $\mathcal{T}^{\boldsymbol{\pi}}$ repeatedly to an initial function $Q_i^{(0)}$.

\begin{proposition}
(Variational Policy Evaluation). For fixed $\boldsymbol{\pi}$ and the variational distribution $q$, consider the modified Bellman operator $\mathcal{T}^{\boldsymbol{\pi}}$ in (\ref{eq:bellmanMain}) and an arbitrary initial function $Q_i^{(0)}:\mathcal{S}\times\mathcal{A}\rightarrow \mathbb{R}$, and define $Q_i^{(k+1)}=\mathcal{T}^{\boldsymbol{\pi}}Q_i^{(k)}$. Then, $Q_i^{(k)}$ converges to $Q_i^{\boldsymbol{\pi}}$ defined in (\ref{eq:value_function11}).
\end{proposition}

{\it{Proof}}. See Appendix B.
\vspace{0.3em}

In the policy improvement step, we update the policy and the variational distribution by using the value function evaluated in the policy evaluation step. Here,  each agent updates its policy and variational distribution while keeping other agents' policies fixed as follows: {\small$(\pi^i_{k+1}, q_{k+1}) =$
\begin{align}
    &\mathop{\arg\max}_{\pi^i, q} \mathbb{E}_{\scriptsize\begin{array}{c}(a^i,a^{-i})\sim (\pi^i, \pi_{k}^{-i})\\
    z_k \sim P_Z
    \end{array}} \Bigg[Q_i^{\boldsymbol{\pi}_k}(s,\boldsymbol{a})
      -\beta \log \tilde{\pi}^i(a^i|s)  \nonumber \\ & ~~~~~~~~~~~~~~~~~~~~~~~~~ + \frac{\beta}{N} \sum_{j\neq i}\log q^{(i,j)}(a^i,a^j|s) ) \Bigg], \label{eq:policy_improvementMain}
\end{align}
}where $a^{-i} \triangleq \{a^1,\cdots,a^N\} \backslash \{a^i\}$ and $\pi_k^{-i}$ is the collection the policies for all agents except Agent $i$ at the $k$-th iteration. Then, we have the following proposition regarding the improvement step.

\begin{proposition}
(Variational Policy Improvement). Let $\pi_{new}^i$ and $q_{new}$ be the updated policy and the variational distribution from (\ref{eq:policy_improvementMain}). Then, $Q_i^{\pi^i_{new}, \pi^{-i}_{old}}(s,\boldsymbol{a})\geq Q_i^{\pi^i_{old}, \pi^{-i}_{old}}(s,\boldsymbol{a})$ for all $(s,\boldsymbol{a}) \in (\mathcal{S}\times \boldsymbol{\mathcal{A}})$.  Here, $Q_i^{\pi^i_{new}, \pi^{-i}_{old}}(s,\boldsymbol{a})$ means $Q_i^{\boldsymbol{\pi}}(s,\boldsymbol{a})|_{\boldsymbol{\pi}=(\pi^i_{new}, \pi^{-i}_{old})}$.
\end{proposition}
{\it{Proof}}.  See Appendix B.
\vspace{0.3em}

The modified policy iteration is defined as applying the variational policy evaluation and variational improvement steps in an alternating manner.
Each agent trains its policy, critic and the variational distribution to maximize its  objective function (\ref{eq:obj_mmi_bound_agent}).


\section{Algorithm Construction}
\label{sec:BasicSetup}

\begin{figure}[t]
\begin{center}
\begin{tabular}{c}
     \includegraphics[width=0.45\textwidth]{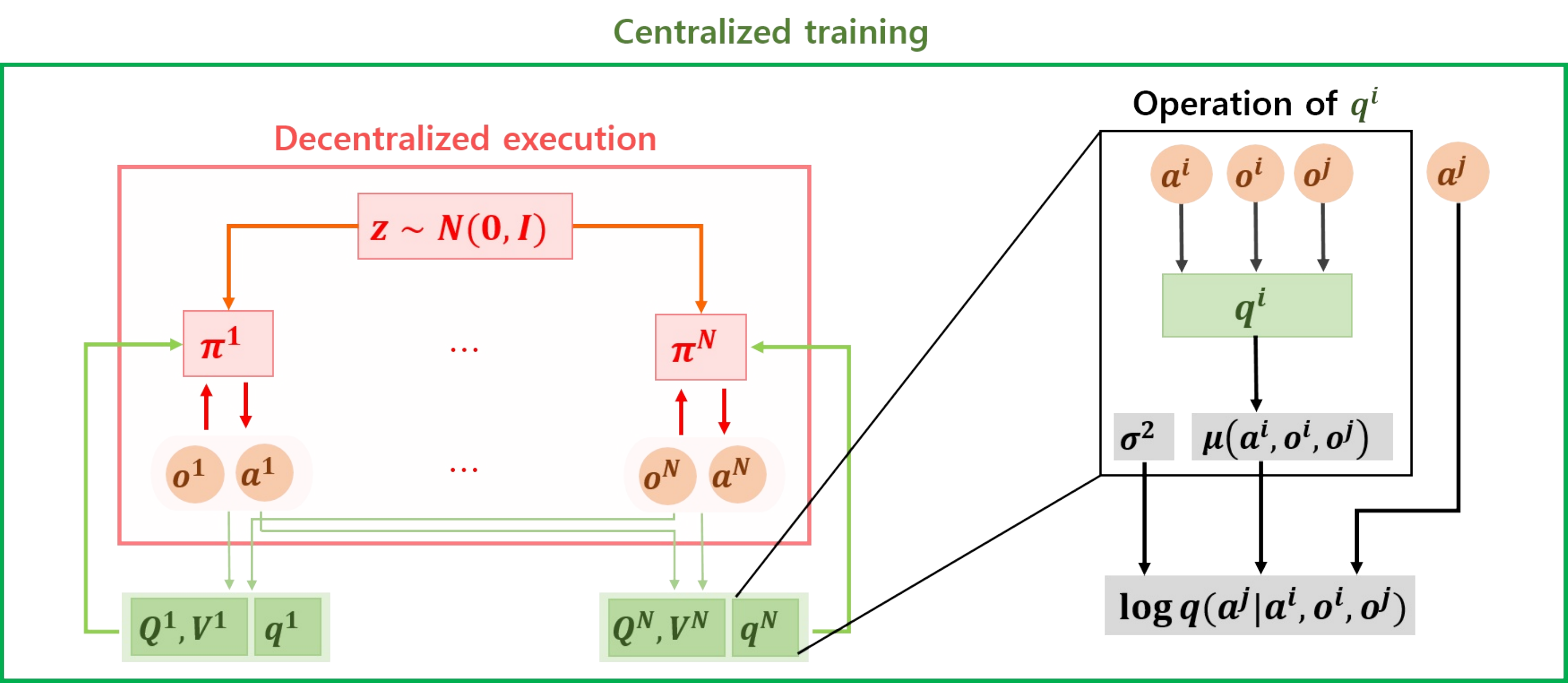}
\end{tabular}
\caption{Overall operation of the proposed VM3-AC. We only need the operation in the red box after training.}
\label{fig:overview}
\end{center}
\end{figure}

Summarizing the development above, we now propose the variational maximum mutual information multi-agent actor-critic (VM3-AC) algorithm, which can be applied to  continuous and partially observable multi-agent environments under CTDE.
The overall operation of VM3-AC is shown in Fig. \ref{fig:overview}.
Under CTDE, each agent's policy is conditioned only on local observation, and centralized critics are conditioned on either the environment state or the observations of all agents, depending on the situation \cite{2017Lowe}.
Let $\boldsymbol{x}$ denote either
the environment state $s$ or the observations  of all agents $(o_1,\cdots,o_N)$, whichever is used.
In order to  deal with  the large continuous state-action spaces, we adopt deep neural networks to approximate the required functions.
For Agent $i$, we parameterize 
 the policy  as $\pi_{\phi^i}^i(a|o^i,z)$ with parameter $\phi^i$, 
the variational distribution  as $q_{\xi^i}(a^j|a^i,(o^i,o^j))$ with parameter $\xi^i$,  
the state-value function  as $V^i_{\psi_i}(\boldsymbol{x})$ with parameter $\psi^i$, and
two action-value functions  as $Q^i_{\theta^{i,1}}(\boldsymbol{x},\boldsymbol{a})$ and $Q^i_{\theta^{i,2}}(\boldsymbol{x},\boldsymbol{a})$ with parameters $\theta^{i,1}$ and $\theta^{i,2}$.
Note that in the original variational distribution, $a_t^j$ is conditioned on $a_t^i$ and $s_t$. In the partially observable case, we replace $s_t$ with $(o_i,o_j)$.

For the prior distribution $P_Z$ of the injection variable $z_t$, we use  zero-mean multivariate Gaussian distribution with identity covariance matrix, i.e., $z_t\sim \mathcal{N}(\mathbf{0},\mathbf{I})$, where the dimension is a hyperparameter, given in Appendix E. We further assume that the class $\mathcal{Q}$ of the variational distribution is multivariate Gaussian distribution with constant covariance matrix $\sigma^2\mathbf{I}$ with dimension of the action dimension, i.e., $\mathcal{Q}=\{q_{\xi^i}(a^j|a^i,(o^i, o^j))=\mathcal{N}(\mu_{\xi^i}(a^i,o^i, o^j), \sigma^2\mathbf{I})\}$, where $\mu_{\xi^i}(a^i,o^i, o^j)$ is the mean of the distribution.

\textbf{Centralized Training}  The parameterized value functions, the policy, and the variational distribution are trained based on proper loss functions derived from Section \ref{subsec:ModifiedPI} in a similar way to the training in SAC in a centralized manner.  During the centralized training, correlated exploration works so as to find a good set of joint policies of the $N$ agents due to our common injection variable as  explained in Section 4. Now, we provide the training details and pseudo code.

The value functions $V^i_{\psi_i}(\boldsymbol{x})$, $Q^i_{\theta_i}(\boldsymbol{x},\boldsymbol{a})$ are updated based on the modified Bellman operator defined in (13) and (14). The state-value function $V^i_{\psi_i}(\boldsymbol{x})$ is trained to minimize the following loss function:
\begin{equation}\label{eq:practical_value}
    \mathcal{L}_V(\psi^i)=\mathbb{E}_{s_t\sim D}\left[ \frac{1}{2}(V^i_{\psi^i}(\boldsymbol{x}_t)-\hat{V}^i_{\psi^i}(\boldsymbol{x}_t))^2 \right]
\end{equation}
where $D$ is the replay buffer that stores the transitions $(\boldsymbol{x}_t,\boldsymbol{a}_t, r_t, \boldsymbol{x}_{t+1})$;  $Q^i_{min}(\boldsymbol{x}_t,a_t^i) = \text{min}[Q^i_{\theta^{i,1}}(\boldsymbol{x}_t,a_t^i), Q^i_{\theta^{i,2}}(\boldsymbol{x}_t,a_t^i)]$ is the minimum of the two action-value functions to prevent the overestimation problem \cite{2018Fujimoto}; and 
\begin{align} 
&\hat{V}^i_{\psi^i}(\boldsymbol{x}_t) = \mathbb{E}_{z_t\sim N(0,\boldsymbol{I}),\{a^k\sim \pi^k(\cdot|o^k_t,z_t)\}_{k=1}^{N}}\Bigg[ Q^i_{min}(\boldsymbol{x}_t,\boldsymbol{a}_t) \nonumber \\ & -\beta \log \pi^i_{\phi^i}(a_t^i|o^i_t,z_t)  + \frac{\beta}{N} \sum_{j\neq i}\log q_{\xi^i}^{(i,j)}(a_t^i,a_t^j|o_t^i,o_t^j) \Bigg].
\label{eq:hatViAppend}
\end{align}
Note that in the second term of the RHS of \eqref{eq:hatViAppend}, originally we should have used the marginalized version, $-\beta \log \tilde{\pi}_{\phi^i}^i(a_t^i|o_t^i)=-\beta \log \mathbb{E}_{z_t\sim N(0,\boldsymbol{I})} [\pi_{\phi^i}^i(a_t^i|o_t^i,z_t)]$. However, for simplicity of computation, we took the expectation  $\mathbb{E}_{z_t\sim N(0,\boldsymbol{I})}$ outside the logarithm. Hence, there exists Jensen's inequality type approximation error. We observe that this approximation works well.

The two action-value functions are updated by minimizing the loss
\begin{equation}\label{eq:loss_critic1}
    \mathcal{L}_Q(\theta^i)=\mathbb{E}_{(\boldsymbol{x}_t,\boldsymbol{a}_t)\sim D}\left[ \frac{1}{2}(Q_{\theta^i}(\boldsymbol{x}_t,\boldsymbol{a}_t)-\hat{Q}(\boldsymbol{x}_t,\boldsymbol{a}_t))^2 \right]
\end{equation}
where
\begin{equation}\label{eq:loss_critic2}
    \hat{Q}(\boldsymbol{x}_t,\boldsymbol{a}_t) =r_t(x_t,\boldsymbol{a_t}) + \gamma \mathbb{E}_{\boldsymbol{x}_{t+1}}[V_{\overline{\psi}^i}\boldsymbol({x}_{t+1})]
\end{equation}
and $V_{\overline{\psi}^i}$ is the target value network, which is updated by the exponential moving average method. We implement the reparameterization trick to estimate the stochastic gradient of policy loss. Then, the action of agent $i$ is given by $a^i=f_{\phi^i}(s;\epsilon^i,z)$, where $\epsilon^i \sim \mathcal{N}(0,\boldsymbol{I})$ and $z\sim \mathcal{N}(0,\boldsymbol{I})$. The policy for Agent $i$ and the variational distribution are trained to minimize the following policy improvement loss, $\mathcal{L}_{\pi^i, q}(\phi^i, \xi)$
\begin{align}
        &=\mathbb{E}_{{\scriptsize \begin{array}{c}
        s_t\sim D,\\
        \epsilon^i\sim \mathcal{N},\\ z\sim \mathcal{N}
        \end{array}}}\Bigg[ -Q^i_{\theta^{i,1}}(\boldsymbol{x}_t,\boldsymbol{a}) + \beta \log \pi^i_{\phi^i}({a}^i|o^i_t, z) \nonumber \\ &- \frac{\beta}{N} \sum_{j\neq i}\log q_{\xi^i}^{(i,j)}(\pi^i_{\phi^i}({a}^i|o^i_t, z),\pi^j_{\phi^j}({a}^j|o^j_t, z)|o_t^i,o_t^j) \Bigg] \label{eq:loss_actor_va}
\end{align}
where $q_{\xi^i}^{(i,j)}(\pi^i_{\phi^i}({a}^i|o^i_t, z),\pi^j_{\phi^j}({a}^j|o^j_t, z)|o_t^i,o_t^j)$
\begin{align} \label{eq:qxiiij}
 = &\underbrace{q_{\xi^i}(\pi^i_{\phi^i}({a}^i|o^i_t, z)|\pi^j_{\phi^j}({a}^j|o^j_t, z)|o_t^i,o_t^j)}_{(a)} \nonumber \\ & ~~~~ \times  \underbrace{q_{\xi^i}(\pi^j_{\phi^j}({a}^j|o^j_t, z)|\pi^i_{\phi^i}({a}^i|o^i_t, z)|o_t^i,o_t^j)}_{(b)}.
\end{align}
Again, for simplicity of computation, we took the expectation  $\mathbb{E}_{z_t\sim N(0,\boldsymbol{I})}$ outside the logarithm for the second term in the RHS in \eqref{eq:loss_actor_va}.
Since approximation of the variational distribution is not accurate in the early stage of training and the learning via the term (a) in  \eqref{eq:qxiiij} is more susceptible to approximation error, we propagate the gradient only through the term (b)  in  \eqref{eq:qxiiij} to  make learning stable.
Note that minimizing $-\log q_{\xi^i}(a^j|a^i,s_t)$ is equivalent to minimizing the mean-squared error between $a^j$ and $\mu_{\xi^i}(a^i,o^i,o^j)$ due to our Gaussian assumption on the variational distribution.

\begin{algorithm}[h]
   \caption{VM3-AC (L=1)}
   \label{alg:VM3-AC}
\begin{algorithmic}
   \STATE \textbf{Centralized training phase}
   \STATE Initialize parameter $\phi^i, \theta^i, \psi^i, \overline{\psi}^i,  \xi^i, ~\forall  i\in \{1,\cdots, N\}$
   \FOR{$episode=1,2,\cdots$}
   \STATE Initialize state $s_0$ and each agent observes $o_0^i$
   \FOR{$t<T$ and $s_t \neq$ terminal}
   \STATE Generate $z_t \sim \mathcal{N}(0,I)$ and select action $a_t^i\sim \pi^i(\cdot|o_t^i,z_t)$ $,\forall i$
   \STATE Execute $\boldsymbol{a_t}$ and each agent $i$ receives $r_t$ and $o_{t+1}^i$
   \STATE Store transitions in $D$
   \ENDFOR
   \FOR{each gradient step}
   \STATE Sample a minibatch from D and generate $z_l \sim \mathcal{N}(0,I)$ for each transition.
   \STATE Update $\theta^i, \psi^i$ by minimizing the loss (\ref{eq:loss_critic1}) and (\ref{eq:loss_critic2})
   \STATE Update $\phi^i, \xi^i$ by minimizing the loss (\ref{eq:loss_actor_va})
   \ENDFOR
   \STATE Update $\overline{\psi}^i$ using the moving average method
   \ENDFOR
   \STATE
   \STATE \textbf{Decentralized execution phase}
   \STATE Initialize state $s_0$ and each agent observes $o_0^i$
   \FOR {each environment step}
   \STATE Select action $a_t^i\sim \pi^i(\cdot|o_t^i,z_t)$ where $z_t=\overrightarrow{0}$ (or sample from the Gaussian random sequence generator with the same seed)
   \STATE Execute $\boldsymbol{a_t}$ and each agent $i$ receives $o_{t+1}^i$
   \ENDFOR
\\
\end{algorithmic}
\end{algorithm}

\textbf{Decentralized Execution}
In the centralized training phase, we pick actions $(a_t^1,\cdots,a_t^N)$  according to $\pi^1(a_t^1|s_t,z_t),\cdots,\pi^N(a_t^N|s_t,\\z_t)$ (or with $s_t$ replaced with $(o_t^1,\cdots,o_t^N)$), where common $z_t$ generated from zero-mean Gaussian distribution is shared under the centralized assumption. However, in the decentralized execution phase, sharing common $z_t$ requires communication among the agents.  To remove this communication necessity, we consider two methods.   First, under the assumption of synchronization, we can make all agents have the same Gaussian random sequence generator and distribute the same seed and initiation timing to this random sequence generator only once in the beginning of the execution phase. In other words, we require all agents to have the same Gaussian random sequence generator and distribute the same seed and initiation timing to these random sequence generators before deployment for the execution phase. (\citet{2019Mahajan2019} also considered that multiple agents share the realization of latent variables in the beginning of the episode.) Second, we exploit the property of zero-mean Gaussian  input variable $z_t$ to the policy network. During the centralized training period, the parameters $\phi^1,\cdots,\phi^N$ of the policy networks $\pi_{\phi^1}^1(a|o^1,z),\cdots,\pi_{\phi^N}^N(a|o^N,z)$ (with input $(o^i,z)$ and output $a$) are learned so that actions $a_t^1,\cdots,a_t^N$ are coordinated for random perturbation input $z_t$ drawn from $P_Z$. Note that the coordination behavior is learned and engraved into the parameters $\phi_1,\cdots,\phi_N$  not into the input $z_t$. So, we only use this stored parameter information during the decentralized execution phase. We apply the common mean value $\mathbb{E}\{z_t\}$ to the $z_t$ input  of the trained policy network  $\pi_{\phi^i}^i(a_t^i|o_t^i,z_t)$ of Agent $i$, $\forall i$. In this case, actions $a_t^1,\cdots, a_t^N$ are independent conditioned on $s_t \ni (o_t^1,\cdots,o_t^N)$  but a specific joint bias (most representative joint bias) is applied to actions  $a_t^1,\cdots, a_t^N$. We expect that this joint bias is helpful and this situation is described in a toy example in Appendix A. In this way,     the proposed algorithm is fully operative under CTDE. The  ablation study is provided in Sec. \ref{sec:Experiment}.

\begin{figure*}[t]
\begin{center}
\begin{tabular}{cccc}
     \includegraphics[width=0.21\textwidth]{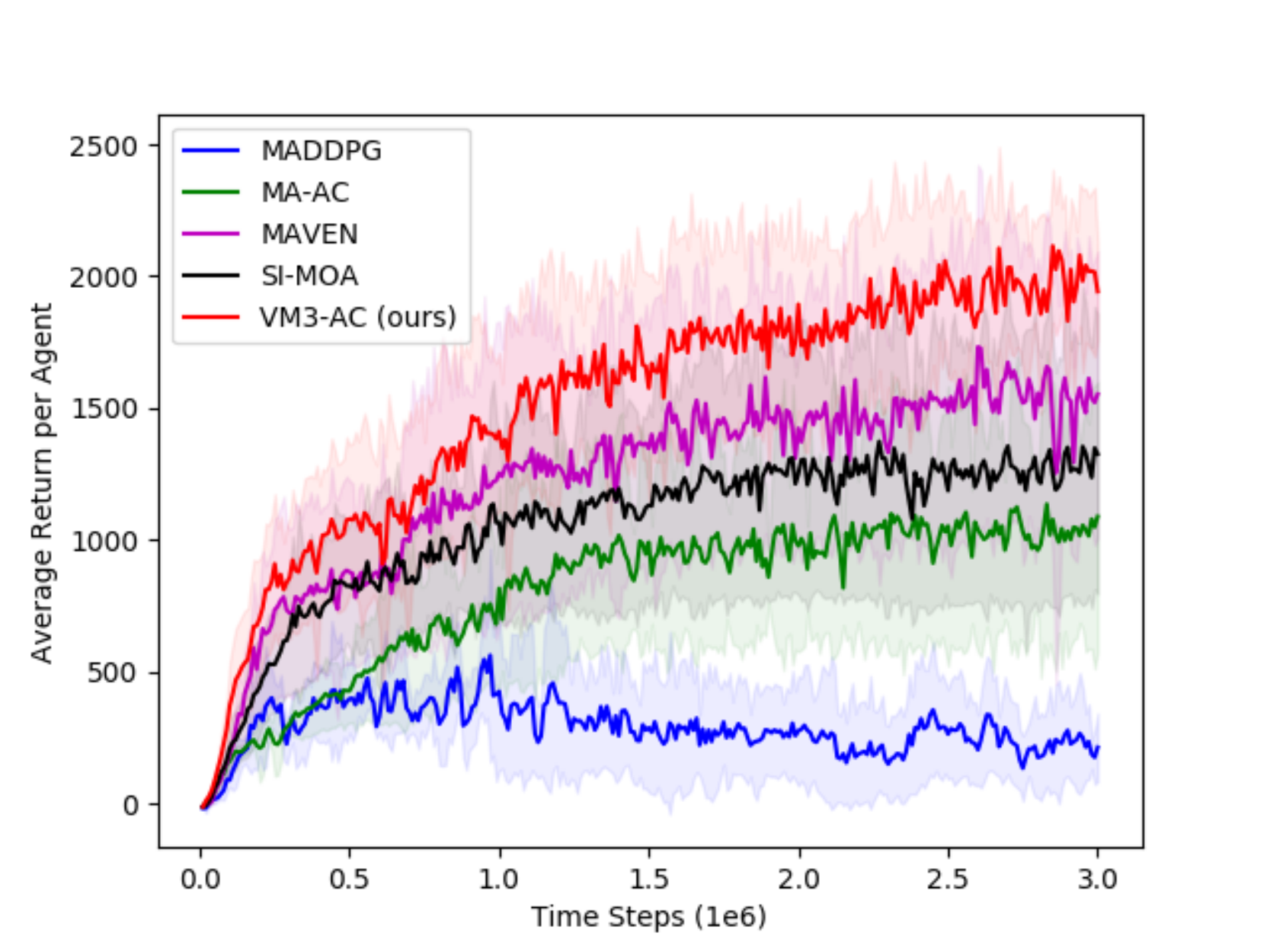} &
     \includegraphics[width=0.21\textwidth]{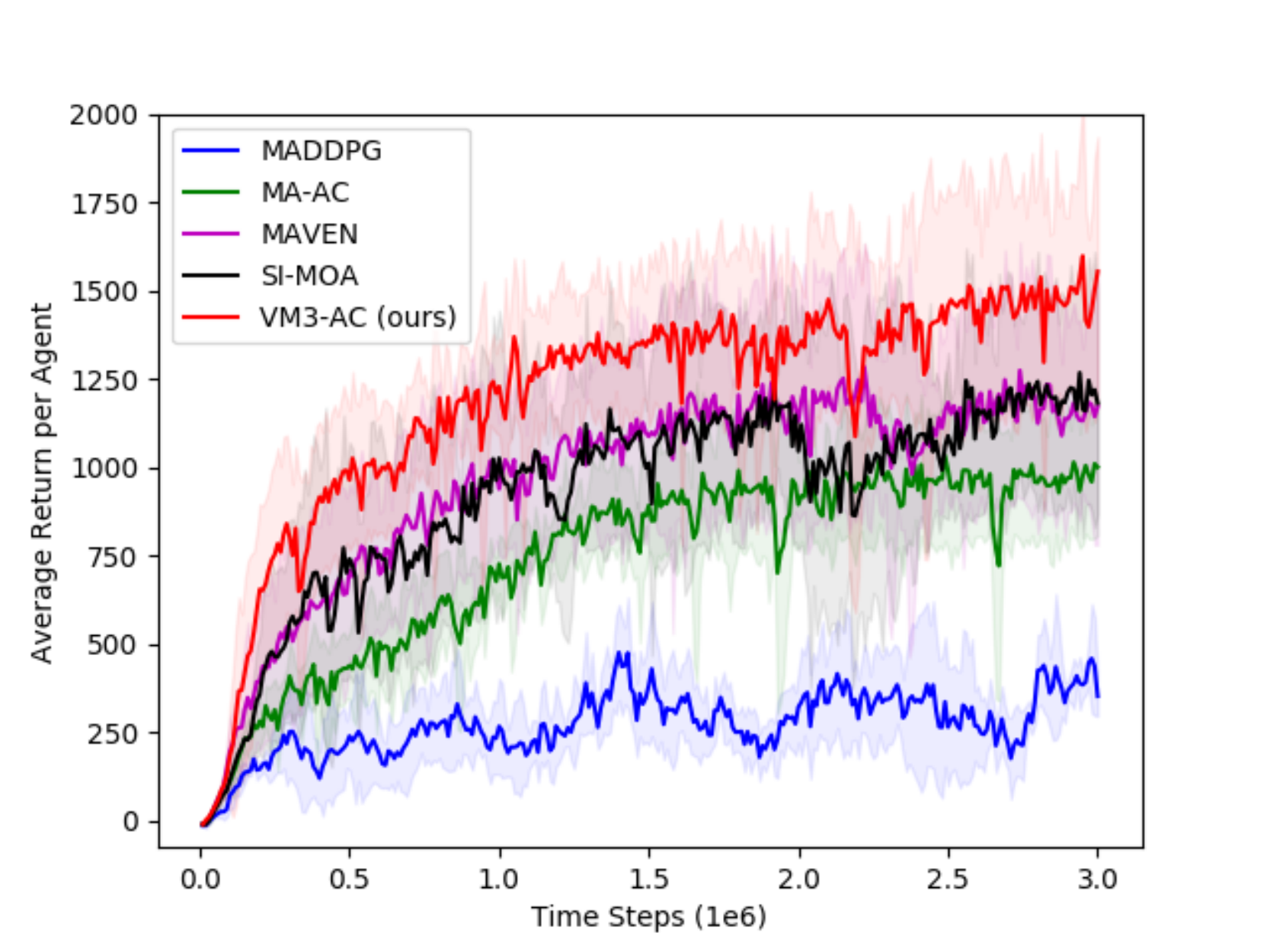} &
     \includegraphics[width=0.21\textwidth]{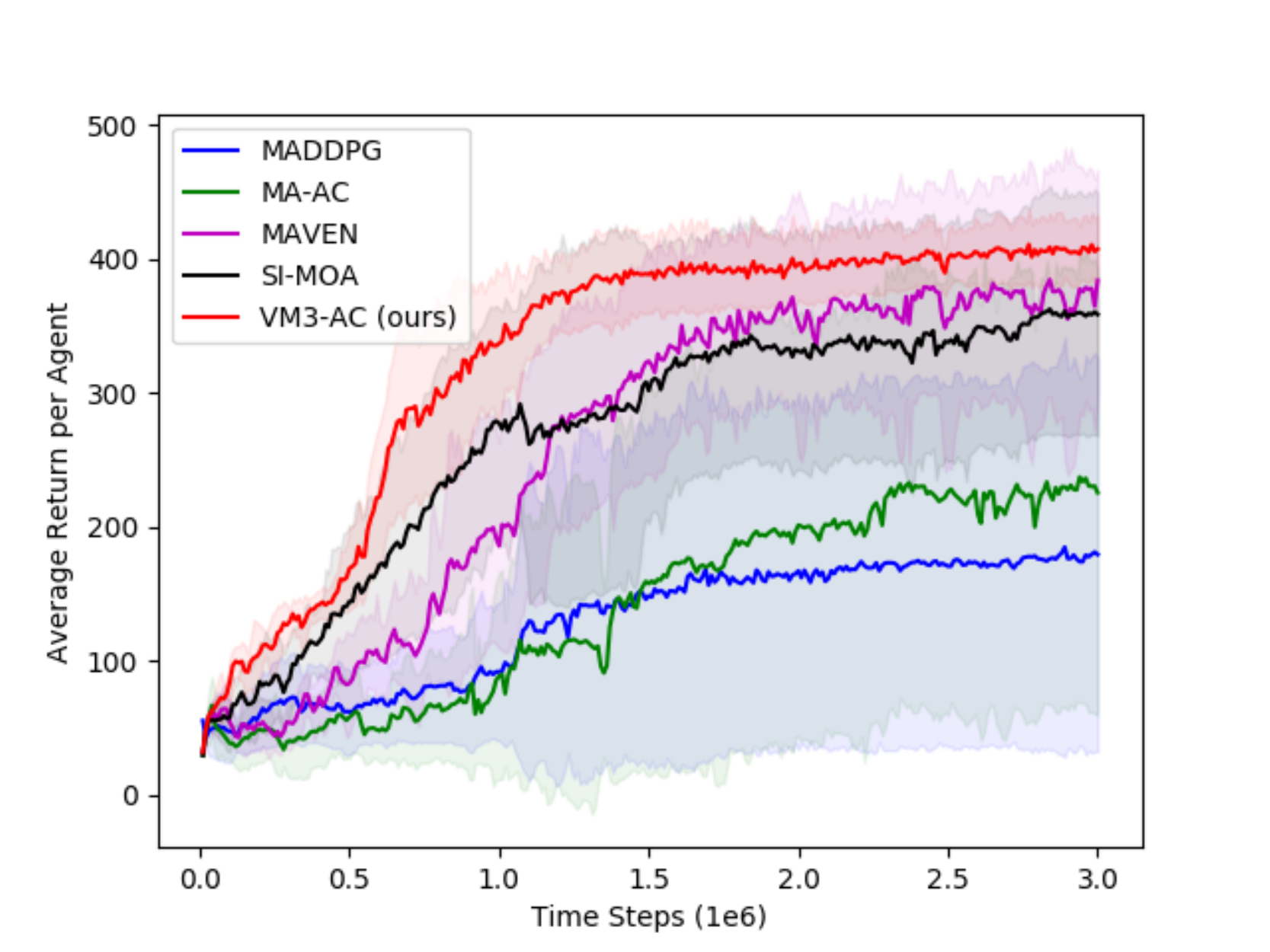} &
     \includegraphics[width=0.21\textwidth]{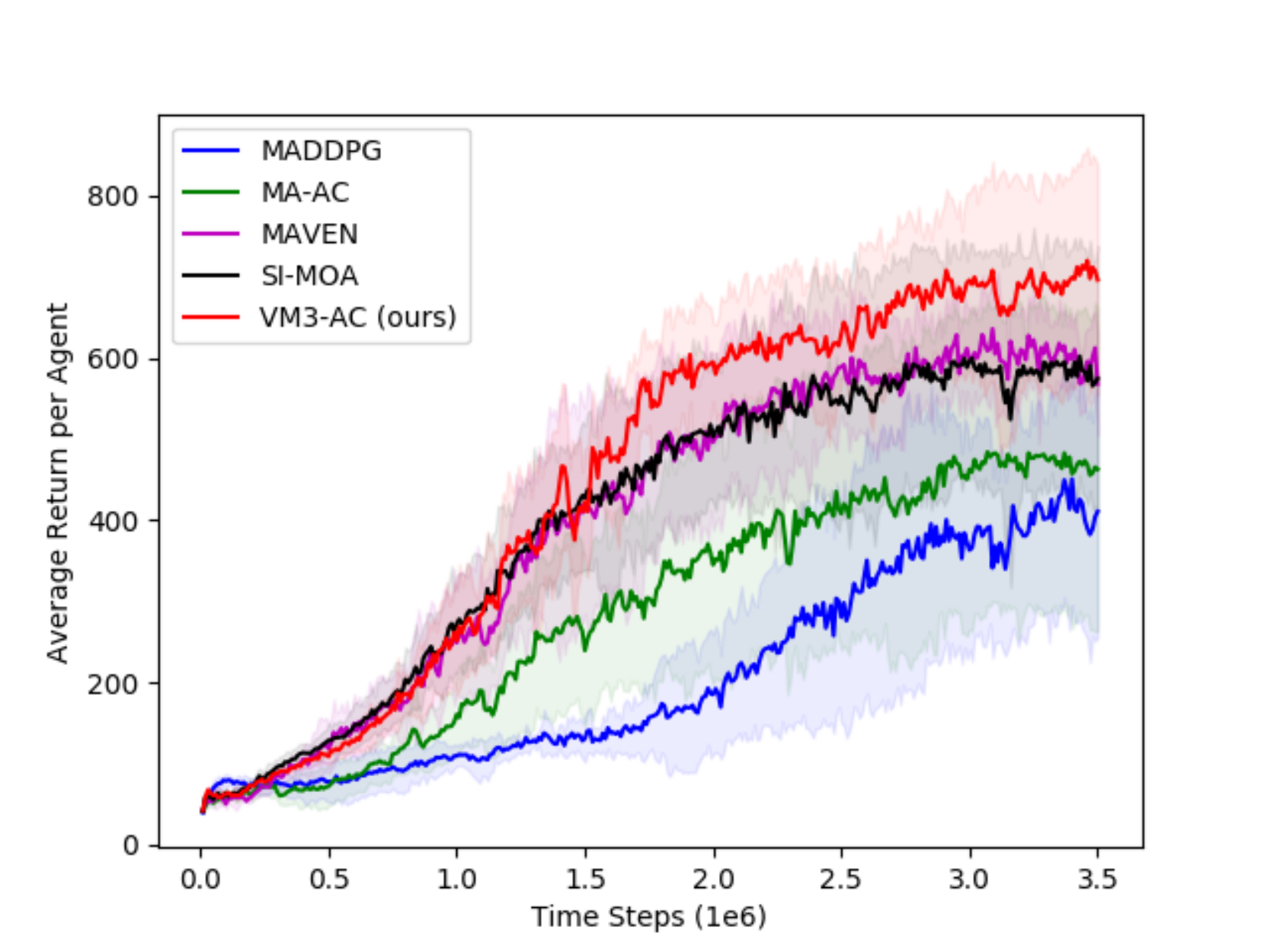} \\
     (a) MW (N=3) & (b) MW (N=4) & (c) PP (N=2) & (d) PP (N=3) \\
     \includegraphics[width=0.21\textwidth]{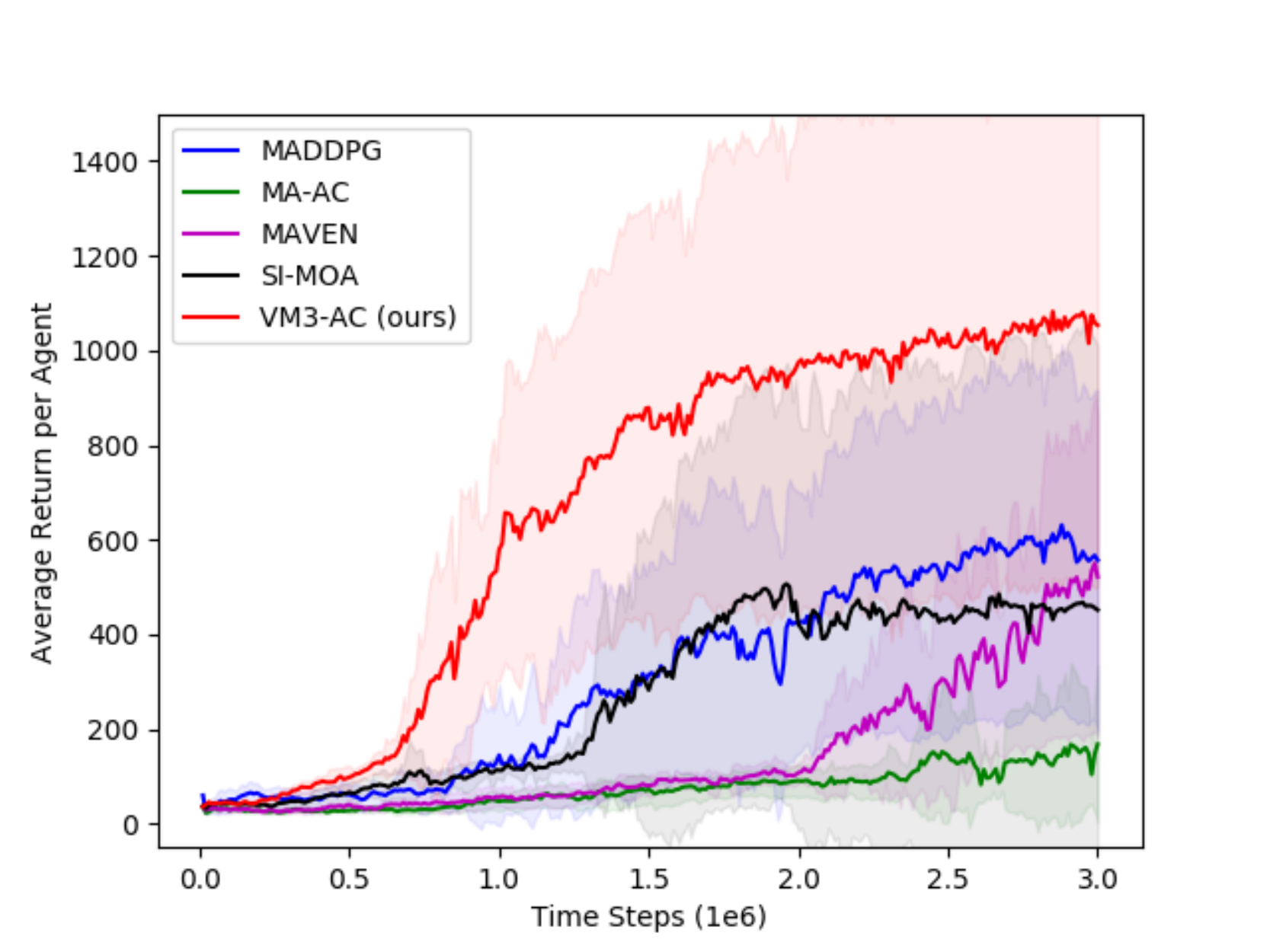} &
     \includegraphics[width=0.21\textwidth]{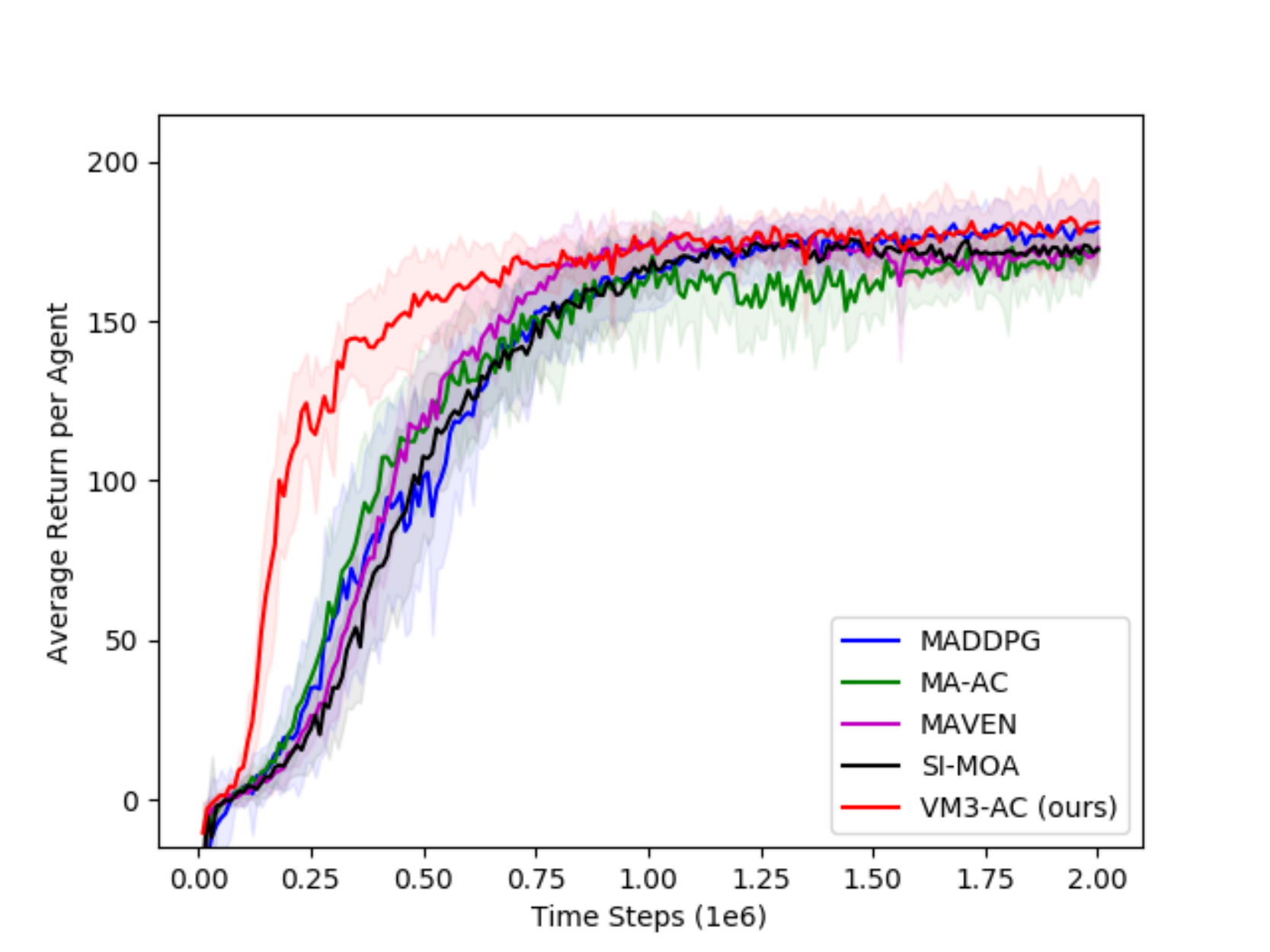} &
     \includegraphics[width=0.21\textwidth]{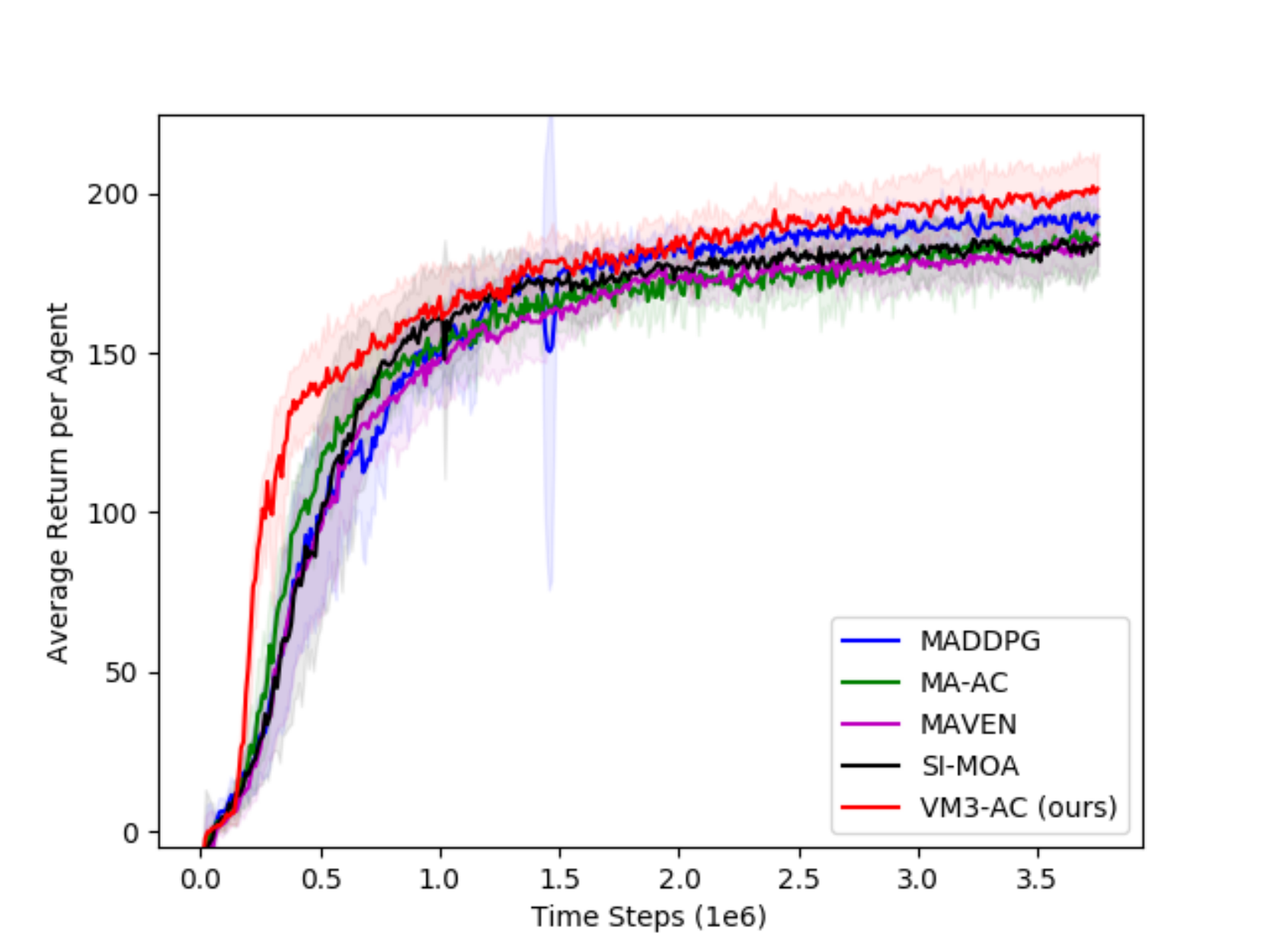} &
     \includegraphics[width=0.21\textwidth]{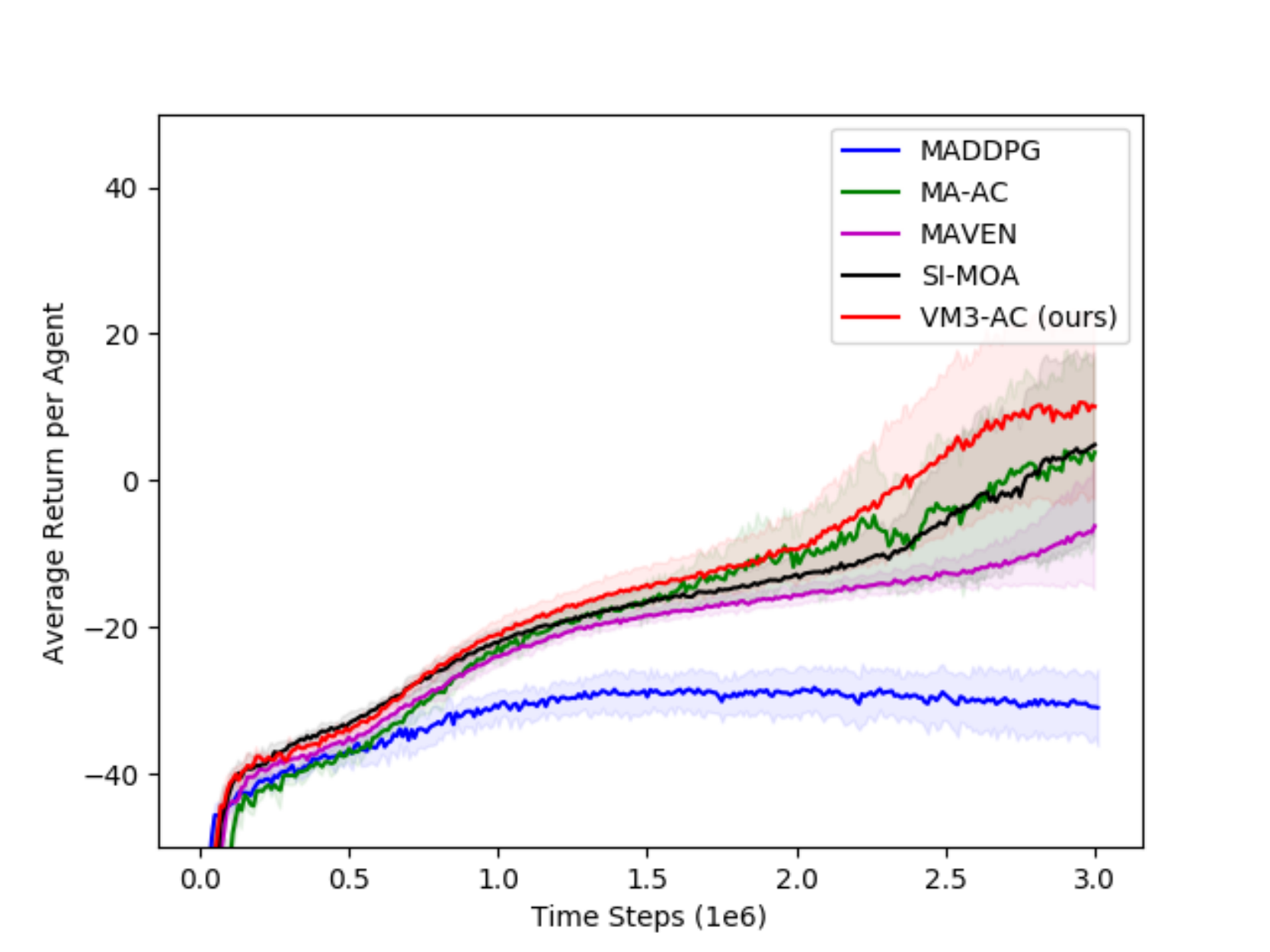}\\
     (e) PP (N=4) & (f) CTC (N=4) & (g) CTC (N=5) & (h) CN (N=3) \\
\end{tabular}
\caption{Performance of MADDPG (blue), MA-AC (green), MAVEN (purple), SI-MOA (black), and VM3-AC (the proposed method, red) on multi-walker environments (a)-(b), predator-prey (c)-(e), cooperative treasure collection (f)-(g), and cooperative navigation (f). (MW, PP, CTC, and CN denote multi-walker, predator-prey, cooperative treasure collection and cooperative navigation, respectively)}
\label{fig:results}
\end{center}
\end{figure*}

\section{Experiment}
\label{sec:Experiment}

In this section, we provide numerical results on both continuous and discrete action tasks.

\textbf{Experiment on continuous action tasks} ~~ We consider the following continuous action tasks with the varying number of agents: multi-walker \cite{2018Gupta}, predator-prey \cite{2017Lowe}, cooperative treasure collection \cite{maac}, and cooperative navigation \cite{2017Lowe}. The detailed setting of each task is provided in Appendix F. 
Here, we considered four baselines: 1) MADDPG \cite{2017Lowe} - an extension of DDPG with a centralized critic to train a decentralized policy for each agent. 2) Multi-agent actor-critic (MA-AC) - a variant of VM3-AC ($\beta=0)$ without the latent variable. 3) Multi-agent variational exploration (MAVEN) \cite{2019Mahajan2019}. Similarly to VM3-AC, MAVEN introduced  latent variable and variational approach for optimizing the mutual information. However, MAVEN does not consider the mutual information between actions but considers the mutual information between the latent variable and trajectories of the agents.
4) Social Influence with MOA (SI-MOA)  \cite{2019Jaques}, which is explained in Section \ref{sec:Background}.  Both MAVEN and SI-MOA are implemented on top of MA-AC since we consider continuous action-space environments.


Fig. \ref{fig:results} shows the learning curves for the considered four environments with the different numbers of agents. The y-axis denotes the average of all agents' rewards averaged over 7 random seeds, and the x-axis denotes the time step. The hyperparameters including the temperature parameter $\beta$ and the dimension of the latent variable are provided in Appendix E.  As shown in Fig. \ref{fig:results}, VM3-AC outperforms the baselines in the considered environments. Especially, in the case of the multi-walker environment, VM3-AC has a large performance gain over existing state-of-the-art algorithms. This is because the agents in the multi-walker environment are strongly required to learn simultaneous coordination in order to obtain high rewards. In addition, the agents in the predator-prey environment, where the number of agents is four, should spread out in groups of two to get more rewards. In this environment, VM3-AC also has a large performance gain. Thus, it is seen that  the proposed MMI framework improves  performance in complex multi-agent tasks requiring high-quality coordination. It is observed that both MAVEN and SI-MOA outperform the basic algorithm MA-AC but not VM3-AC. Hence, the numerical results  show that the way of using MI by the proposed VM3-AC algorithm has some advantages over those by MAVEN and SI-MOA, especially for MARL tasks requiring coordination of concurrent actions.

\begin{figure*}[t]
\begin{center}
\begin{tabular}{ccc}
    \includegraphics[width=0.25\textwidth]{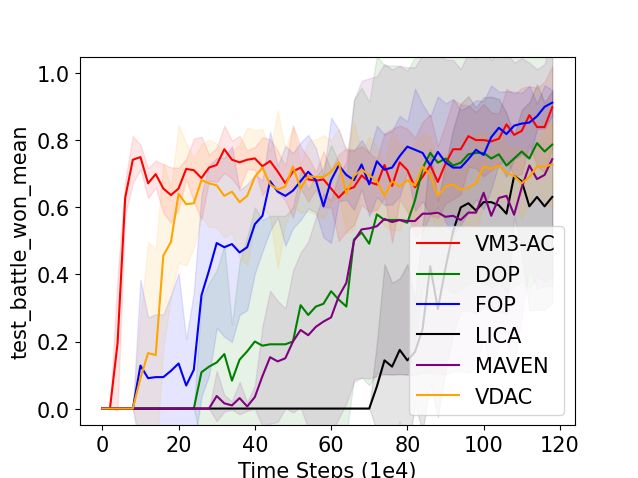} &
     \includegraphics[width=0.25\textwidth]{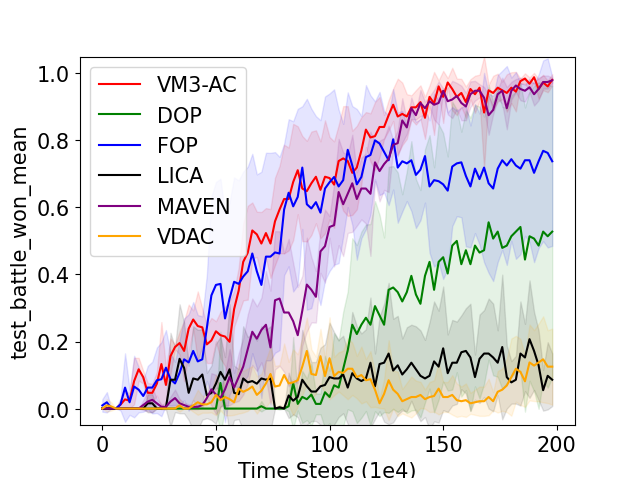} &
     \includegraphics[width=0.25\textwidth]{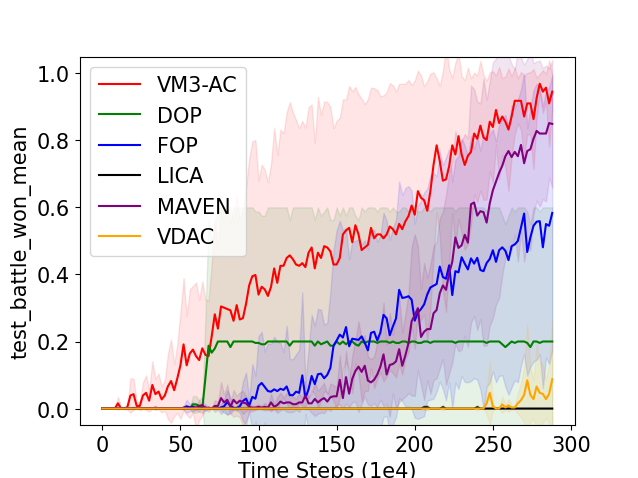} \\
     (a) \textit{3m} & (b) \textit{2s3z} & (c) \textit{3s vs 3z}\\
\end{tabular}
\caption{Performance of DOP (green), FOP (blue), LICA (black), MAVEN (purple), VDAC (orange) and VM3-AC (red) on three maps in the modified SMAC environment.}
\label{fig:result_smac}
\end{center}
\end{figure*}

\textbf{Experiment on discrete action task} ~~ We also considered the StarcraftII micromanagement benchmark (SMAC) environment \cite{SMAC}. We modified the SMAC environment to be sparse by giving rewards when an ally or an enemy dies and a time penalty. Thus, in the case of \textit{3s vs 3z}, where we need to control three stalkers to beat the three zealots (enemy), the reward is hardly obtained because it takes a long time to remove a zealot. We provided the detailed setting of the modified SMAC environment in Appendix G. We considered five state-of-the-art baselines: DOP \cite{DOP}, FOP \cite{FOP}, LICA \cite{LICA}, MAVEN \cite{2019Mahajan2019}, and VDAC \cite{VDAC}. We implemented VM3-AC on the top of FOP by introducing the latent variable and replacing the entropy term in \cite{FOP} with the MI. Fig. \ref{fig:result_smac} shows the performances of VM3-AC and the baselines on three maps in SMAC. It is observed that VM3-AC outperforms the baselines. Especially on 3svs3z, in which reward is highly sparse, VM3-AC outperforms the baselines in terms of both training speed and final performance.

\subsection{Ablation Study and Discussion}\label{sec:ablation}

In this subsection, we provide ablation studies and discussion on the major techniques and hyperparameters of VM3-AC: 1) mutual information versus entropy 2) the latent variable, 3) the temperature parameter $\beta$, 4) injecting zero vector instead of the latent variable $z$ to policies in the execution phase and 5) scalability.

\textbf{Mutual information versus entropy:} The proposed MI framework maximizes the sum of the action entropy and the negative of the cross entropy of the variational conditional distribution relative to the true conditional distribution, which provides a lower bound of MI between actions. As aforementioned, maximizing the sum of the action entropy and the negative of the cross entropy of the variational conditional distribution relative to the true conditional distribution enhances exploration and  predictability for other agents' actions. Hence, the proposed MI framework enhances correlated exploration among agents.

\begin{figure*}[t]
\begin{center}
\begin{tabular}{cccc}
     \includegraphics[width=0.2\textwidth]{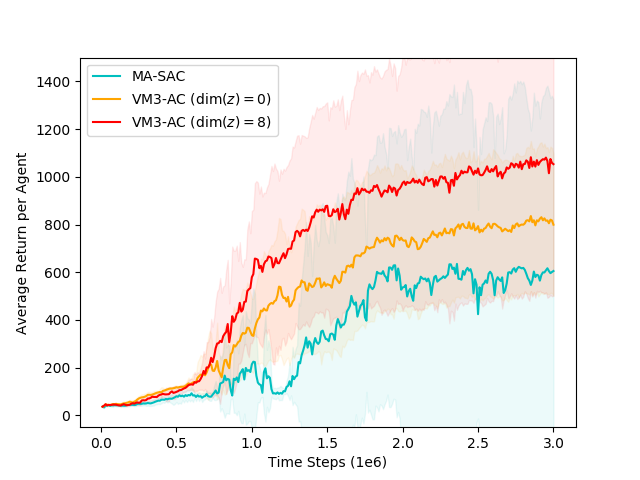} &
     \includegraphics[width=0.2\textwidth]{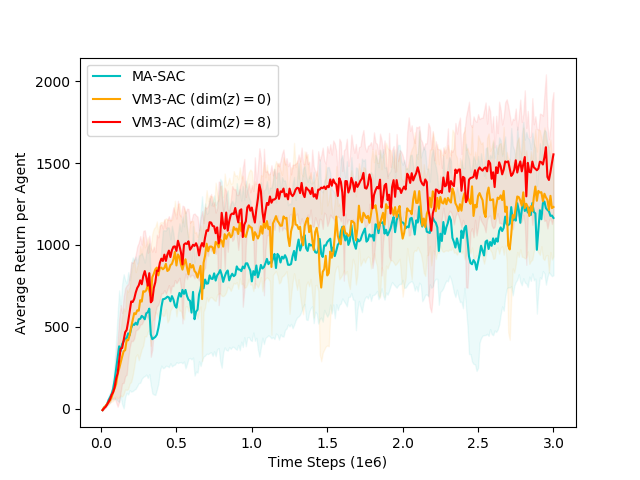} &
     \includegraphics[width=0.2\textwidth]{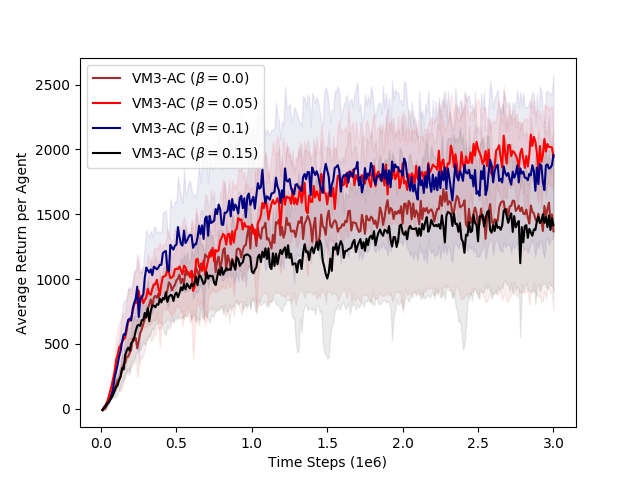} &
     \includegraphics[width=0.2\textwidth]{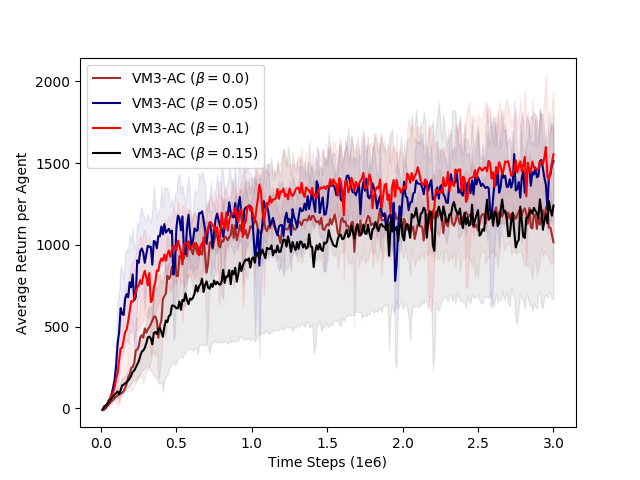} \\
     (a) PP (N=4) & (b) MW (N=4) & (c) MW (N=3) & (d) MW (N=4)
\end{tabular}
\caption{(a) and (b): VM3-AC (red), VM3-AC without latent variable (orange), and MA-SAC (cyan) and (c) and (d): performance with respect to the temperature parameter}
\label{fig:ablations}
\end{center}
\end{figure*}

\begin{figure}[h]
\begin{center}
\begin{tabular}{c}
     \includegraphics[width=0.45\textwidth]{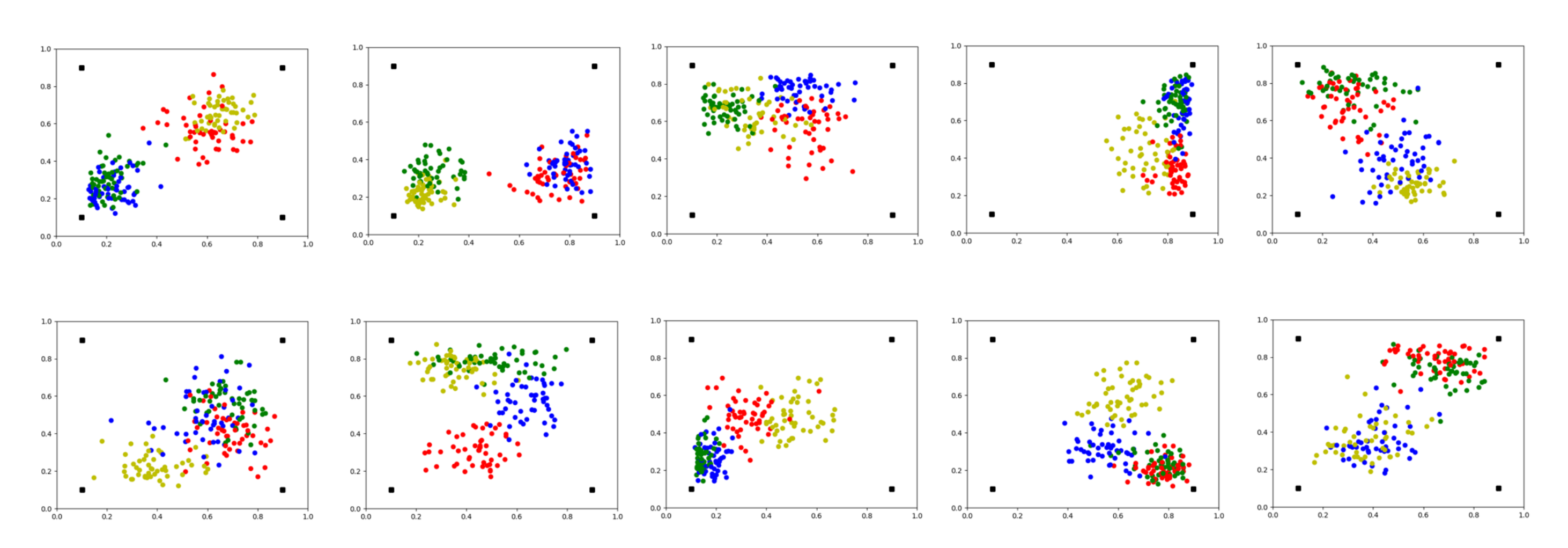}
\end{tabular}
\caption{The positions of four agents after five time-steps after the episode begins in the early stage of the training:  1st row - VM3-AC  and 2nd row - MA-SAC. The figures in column correspond to a different seed. The black squares are the preys and each color except black shows the position of each agent.}
\label{fig:mi_sac}
\end{center}
\vspace{-3ex}
\end{figure}

We compared VM3-AC with multi-agent-SAC (MA-SAC), which is an extension of maximum entropy soft actor-critic (SAC) \cite{2018Haarnoja}  to multi-agent setting.  For MA-SAC, we extended SAC to multi-agent settings in the manner of independent learning. Each agent trains its decentralized policy using decentralized critic to maximize the weighted sum of the cumulative return and the entropy of its policy. Adopting the framework of CTDE, we replaced decentralized critic with centralized critic which incorporates observations and actions of all agents.

We performed an experiment in the predator-prey environment with four agents where the number of required agents to catch the prey is two. In this environment, the agents started at the center of the map. Hence, the agents should spread out in the group of two to catch  preys efficiently. Fig.\ref{fig:mi_sac} shows the positions of the four agents at five time-steps after the episode starts. The first and second rows in Fig.\ref{fig:mi_sac} show the results of VM3-AC and  MA-SAC in the early stage of the training, respectively. It is seen that the agents of VM3-AC explore in the group of two while the agents of MA-SAC tend to explore independently. We provided the performance comparisons of VM3-AC with MA-SAC in Fig.\ref{fig:ablations} (a) and (b).

\textbf{Latent variable:} The role of the latent variable is to induce MI among concurrent actions and inject an additional degree of freedom for action control. We compared VM3-AC and VM3-AC without the latent variable (implemented by setting $\mbox{dimension}(z_t)=0$) in the multi-walker environment. In both cases, VM3-AC yields better performance than VM3-AC without the latent variable as shown in Fig.\ref{fig:ablations}(a) and  \ref{fig:ablations}(b). Here, the gain by VM3-AC without the latent variable (i.e., $\mbox{dimension}(z_t)=0$) over MA-SAC is solely due to passive modeling $p(a_t^j|a_t^i,s_t)$ by using $q(a_t^j|a_t^i,s_t)$, not including active injection of coordination by $z_t$.

\textbf{Injecting mean vector $\mathbb{E}\{z_t\}$ to  the $z_t$-input of policy network $\pi^i_{\phi^i}(\cdot|o_t^i,z_t)$ during the execution phase:} As mentioned in Section \ref{sec:BasicSetup}, we applied the mean vector of $z_t$, i.e., $\mathbb{E}\{z_t\}$ to the $z_t$-input of the policy deep neural network $\pi^i_{\phi^i}(\cdot|o_t^i,z_t)$  during the execution phase so as to execute actions without communication in the execution phase. We compared the performance of decentralized policies that use the mean vector $\mathbb{E}\{z_t\}$ and decentralized policies which use the latent variable $z_t$ assuming communication.
We used deterministic evaluation based on 20 episodes generated by the corresponding deterministic policy, i.e., each agent selects action using the mean network of Gaussian policy $\pi_{\phi^i}^i$. We averaged the return over 7 seeds, and the result is shown in Table \ref{table:ablation}. It is seen that the mean vector replacement method yields almost the same performance and enables fully decentralized execution without noticeable performance loss. Please see Appendix A for intuition.

\begin{table}[h!]
\caption{Impact of replacing the latent variable $z_t\sim \mathcal{N}(0,\mathbf{I})$ with mean vector $z_t = \mathbb{E}(z_t)$ in the execution phase}
\centering
\begin{tabular} { |c | c c c |} 
 \hline
& PP (N=2) & PP (N=3) & PP (N=4) \\
 \hline\hline
 $z_t\sim \mathcal{N}(0,\mathbf{I})$ & $413$ & $734$ & $1123$ \\ 
$z_t = \mathbb{E}(z_t)$ & $409$ & $743$ & $1147$ \\
 \hline
\end{tabular}
\label{table:ablation}
\end{table}

\textbf{Temperature parameter $\beta$:} The role of temperature parameter $\beta$ is to control the relative importance between the reward and the MI. We evaluated VM3-AC by varying  $\beta=[0, 0.05, 0.1, 0.15]$ in the multi-walker environment with $N=3$ and $N=4$. Fig. \ref{fig:ablations}(c) and \ref{fig:ablations}(d) show that VM3-AC with the temperature value around $[0.05, 0.1]$ yields good performance.

\textbf{Scalability:} Many MARL algorithms which use a centralized critic such as MADDPG \cite{2017Lowe} can suffer from the problem of scalability due to increasing joint state-action space as the number of agents increases. VM3-AC can also suffer from the same issue but we can address the problem by adopting an attention mechanism as in MAAC \cite{2018Iqbal}. Additionally, VM3-AC needs more variational approximation networks as the number of agents increases. As many MARL algorithms share the parameters among agents, we can share the parameters for the variational  approximation networks. We expect that parameter sharing can handle the scalability of the proposed method.

\section{Conclusion}
\label{sec:Conclusion}

In this paper, we have proposed a new approach to MI-based coordinated MARL to induce the coordination of concurrent actions under CTDE. In the proposed approach, a common correlation-inducing random variable is injected into each policy network, and  the MI between actions induced by this variable is expressed as a tractable form by using a variational distribution. The derived objective consists of the maximum entropy RL combined predictability enhancement (or uncertainty reduction) for other agents' actions, which can be interpreted as correlated exploration. We evaluated the derived algorithm named VM3-AC on both continuous and discrete action tasks and the numerical results show that VM3-AC outperforms other state-of-the-art baselines, especially in multi-agent tasks requiring high-quality coordination among agents.

\textbf{Limitation} ~~~ One can think sharing the common variable requires communication between agents. To handle this, we introduced two methods including sharing a Gaussian random sequence generator at the beginning of the episode and injecting the mean vector into the latent vector in the execution. Here, reference timing information on top of time step synchronization is required for the method of sharing a Gaussian random generator. This requirement of communication is one limitation of our work, but we provided an ablation study on this alternative and it was seen that the alternative performs well. 

\textbf{Future Work} Communication is also a promising approach to enhance coordination between agents \cite{kim2021communication}. We believe our mutual information framework combined with communication-based learning in MARL has the potential to yield significant benefits. We leave it as a future work.

\section{Acknowledgements}
This work was partly supported  by Institute for Information \& communications Technology Planning \& Evaluation(IITP) grant funded by the Korea government(MSIT) (No. 2022-0-00469) and the National Research Foundation of Korea(NRF) grant funded by the Korea government(MSIT). (NRF-2021R1A2C2009143)

\bibliographystyle{ACM-Reference-Format} 
\bibliography{main}

\onecolumn

\section*{Appendix A: Correlation Based on Common $Z_t$ and Basic Idea}
\label{sec:CorrelationExample}

\begin{figure}[h]
\begin{center}
\begin{tabular}{c}
     \includegraphics[width=0.45\textwidth]{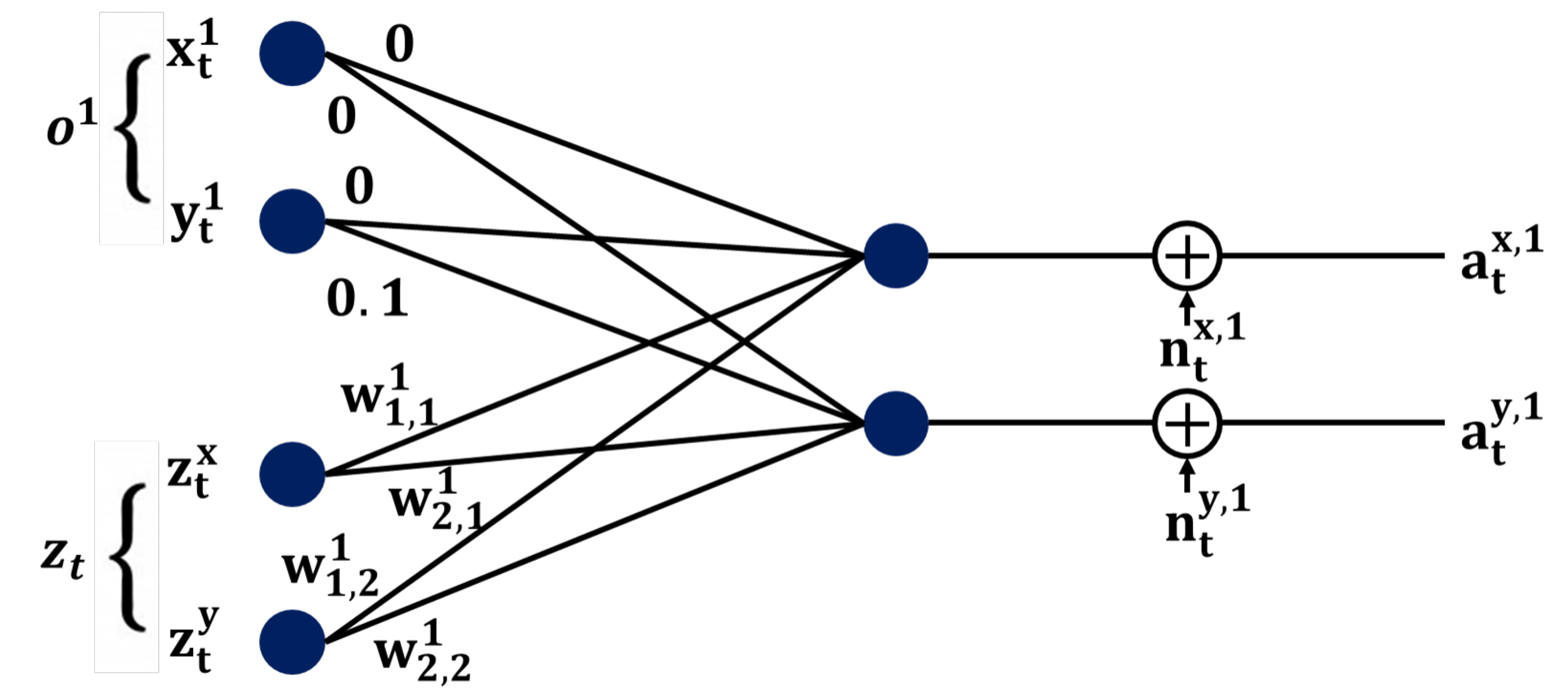}
\end{tabular}
\caption{Graphical representation of the policy function of Agent 1,  eq. (\ref{eq:append_toy_a1})}
\label{fig:ToyExampleNetDiagram}
\end{center}
\end{figure}

Here, we provide a toy example explaining our idea. The example is as follows.  We have  two agents: Agents $1$ and $2$ in a 2-dimensional half-plane $(x,y)$ with $y >0$. The state is the locations of the two agents, i.e., $s_t=((x_t^1,y_t^1),(x_t^2,y_t^2))$, where $(x_t^i,y_t^i)$ is the location of Agent $i$.
The action of each agent is the displacement, i.e., the action of Agent $i$ is $a_t^i=(\Delta x_t^i,\Delta y_t^i)$, $i=1,2$.  The location of Agent $i$ at time $t+1$ is determined as a function of the  state and action at current time $t$: 
\[
(x_{t+1}^i,y_{t+1}^i)=(x_t^i,y_t^i)+(\Delta x_t^i,\Delta y_t^i).
\]
Suppose that Agent $i$ can only observe its own location $o^i=(x_t^i,y_t^i)$ and suppose that 22
  the policies  $\pi^1(a_t^1|o_t^1,z_t)$ and   $\pi^2(a_t^2|o_t^2,z_t)$ of the two agents are functions of the observation and an additional common random variable $z_t$,  and   given by the following simple linear stochastic model:
\begin{align}
    a_t^1 &=\left[
    \begin{array}{c}
     a_t^{x,1}\\
     a_t^{y,1}
     \end{array}
    \right]=\left[
    \begin{array}{c}
     \Delta x_t^1\\
     \Delta y_t^1
     \end{array}
    \right]
     \nonumber \\ &= \underbrace{\left[
    \begin{array}{cc}
     0&0\\
     0&0.1
     \end{array}
    \right]}_{parameter~for~o_t^1}\underbrace{\left[
    \begin{array}{c}
     x_t^1\\
     y_t^1
     \end{array}
    \right]}_{o_t^1} + \underbrace{\left[
    \begin{array}{cc}
     w_{11}^1&w_{12}^1\\
     w_{21}^1&w_{22}^1
     \end{array}
    \right]}_{parameter~for~z_t~ at~ Agent ~1}\underbrace{\left[
    \begin{array}{c}
     z_t^x\\
     z_t^y
     \end{array}
    \right]}_{z_t}+\underbrace{\left[
    \begin{array}{c}
     n_t^{x,1}\\
     n_t^{y,1}
     \end{array}
    \right]}_{noise~\mathbf{n}_t^1~at~Agent~1} \label{eq:append_toy_a1}\\
    a_t^2 &=\left[
    \begin{array}{c}
     a_t^{x,2}\\
     a_t^{y,2}
     \end{array}
    \right]=\left[
    \begin{array}{c}
     \Delta x_t^2\\
     \Delta y_t^2
     \end{array}
    \right]\nonumber \\ &= \underbrace{\left[
    \begin{array}{cc}
     0&0\\
     0&0.1
     \end{array}
    \right]}_{parameter~for~o_t^2}\underbrace{\left[
    \begin{array}{c}
     x_t^2\\
     y_t^2
     \end{array}
    \right]}_{o_t^2} + \underbrace{\left[
    \begin{array}{cc}
     w_{11}^2&w_{12}^2\\
     w_{21}^2&w_{22}^2
     \end{array}
    \right]}_{parameter~for~z_t~ at~ Agent ~2}\underbrace{\left[
    \begin{array}{c}
     z_t^x\\
     z_t^y
     \end{array}
    \right]}_{z_t}+\underbrace{\left[
    \begin{array}{c}
     n_t^{x,2}\\
     n_t^{y,2}
     \end{array}
    \right]}_{noise~\mathbf{n}_t^2~at~Agent~2},\label{eq:append_toy_a2}
\end{align}
where the two random noise terms $\mathbf{n}_t^1$ and $\mathbf{n}_t^2$ at Agents 1 and 2 are independent random variables; $z_t=\left[
    \begin{array}{c}
     z_t^x\\
     z_t^y
     \end{array}
    \right]$ is a random variable (precisely speaking, random vector) drawn from $P_Z(z)$; and the notation of two consecutive brackets $[\cdot][\cdot]$ means matrix multiplication. Fig. \ref{fig:ToyExampleNetDiagram} describes the policy function of Agent 1 given by  (\ref{eq:append_toy_a1}) in a graphical form. (\eqref{eq:append_toy_a2} can be described in a similar graphical form.)

\vspace{1.5em}

In \eqref{eq:append_toy_a1}, the noise term $\mathbf{n}_t^1$ is added to perturb the action of Agent 1 for exploration around the given term $\left[
    \begin{array}{cc}
     0&0\\
     0&0.1
     \end{array}
    \right]\underbrace{\left[
    \begin{array}{c}
     x_t^1\\
     y_t^1
     \end{array}
    \right]}_{o_t^1}$ for given $s_t$.
In \eqref{eq:append_toy_a2},  the noise term $\mathbf{n}_t^2$ is added to perturb the action of Agent 2 for exploration around the given term $\left[
    \begin{array}{cc}
     0&0\\
     0&0.1
     \end{array}
    \right]\underbrace{\left[
    \begin{array}{c}
     x_t^2\\
     y_t^2
     \end{array}
    \right]}_{o_t^2}$ for given $s_t$. Note that two perturbation terms $\mathbf{n}_t^1$ and $\mathbf{n}_t^2$ are independent. Hence, these two terms induce independent exploration for Agents 1 and 2. That is,  without the $z_t$-induced terms in \eqref{eq:append_toy_a1} and \eqref{eq:append_toy_a2}, $a_t^1$ and $a_t^2$ given $s_t \ni (o_t^1,o_t^2)$ are independent  since in this case only the noise terms $\mathbf{n}_t^1$ and $\mathbf{n}_t^2$ remain and the noise terms are independent random variables by assumption. However, with the $z_t$-induced terms in \eqref{eq:append_toy_a1} and \eqref{eq:append_toy_a2}, $a_t^1$ and $a_t^2$ are correlated and the corresponding covariance matrix is given by
\begin{align}
    \mathbb{E}[ (a_t^1-\mathbb{E}[a_t^1]) (a_t^2-\mathbb{E}[a_t^2])^T|s_t] &= \mathbf{W}^1 \mathbf{C}_z (\mathbf{W}^2)^T,
\end{align}
where $a_t^1$ and $a_t^2$ are column vectors as shown in \eqref{eq:append_toy_a1} and \eqref{eq:append_toy_a2}; $(\cdot)^T$ denotes matrix transpose; $\mathbf{C}_z$ is the covariance matrix of $z_t=\left[
    \begin{array}{c}
     z_t^x\\
     z_t^y
     \end{array}
    \right]$ determined by $p_Z$;  
\[
\mathbf{W}^1 =\left[
    \begin{array}{cc}
     w_{11}^1&w_{12}^1\\
     w_{21}^1&w_{22}^1
     \end{array}
    \right]~~~~~\mbox{and}~~~~~\mathbf{W}^2=\left[
    \begin{array}{cc}
     w_{11}^2&w_{12}^2\\
     w_{21}^2&w_{22}^2
     \end{array}
    \right].
\]
Note that the perturbation structure of $\mathbf{W}^1z_t$ and $\mathbf{W}^2 z_t$  is different from that of $\mathbf{n}_t^1$ and $\mathbf{n}_t^2$. Indeed, we are injecting correlated random perturbation into $a_t^1$ and $a_t^2$ to promote correlation exploration to better explore the joint state-action space. 
By properly designing $\mathbf{C}_z$ (i.e., properly designing $P_Z$), $\mathbf{W}^1$ and $\mathbf{W}^2$, we can impose an arbitrary  correlation structure between $a_t^1$ and $a_t^2$ conditioned on $s_t$.

\begin{figure}[h]
\begin{center}
\begin{tabular}{c}
     \includegraphics[width=0.375\textwidth]{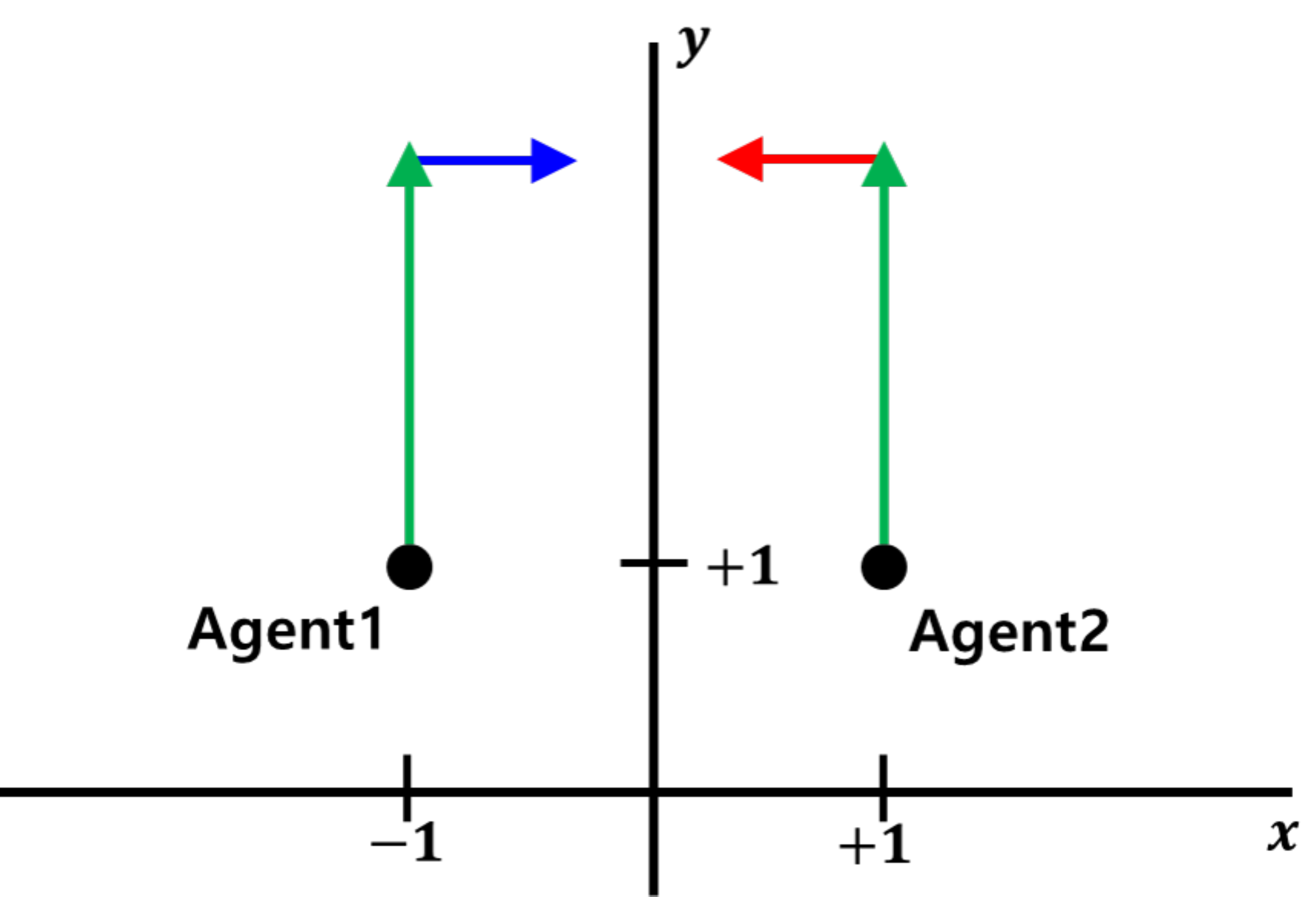}
\end{tabular}
\caption{An example}
\label{fig:ToyExample}
\end{center}
\end{figure}

Now, consider the following joint task. The initial location of Agent 1 is (-1,1) and the initial location of Agent 2 is (1,1). The joint goal is that the two agents meet  while going upward, and an episode ends when the two agents meet, as described in Fig. \ref{fig:ToyExample}.
Suppose that we pick the prior distribution $P_Z(z)$ for $z_t$ as
\begin{align}
    z_t^x &\sim \mbox{Unif}[0,0.1]  \label{eq:ztxDist}\\
    z_t^y &\sim \mbox{Unif}[0,0.1], \label{eq:ztyDist}
\end{align}
where $\mbox{Unif}[a,b]$ means the uniform distribution over interval $[a,b]$. Now, we design the reward 
$r_t(s_t,a_t^1,a_t^2)$ as the distance between the two agents' locations.   We use the policies of the two agents given by \eqref{eq:append_toy_a1} and \eqref{eq:append_toy_a2}. 

\begin{figure}[h]
\begin{center}
\begin{tabular}{cc}
     \includegraphics[width=0.45\textwidth]{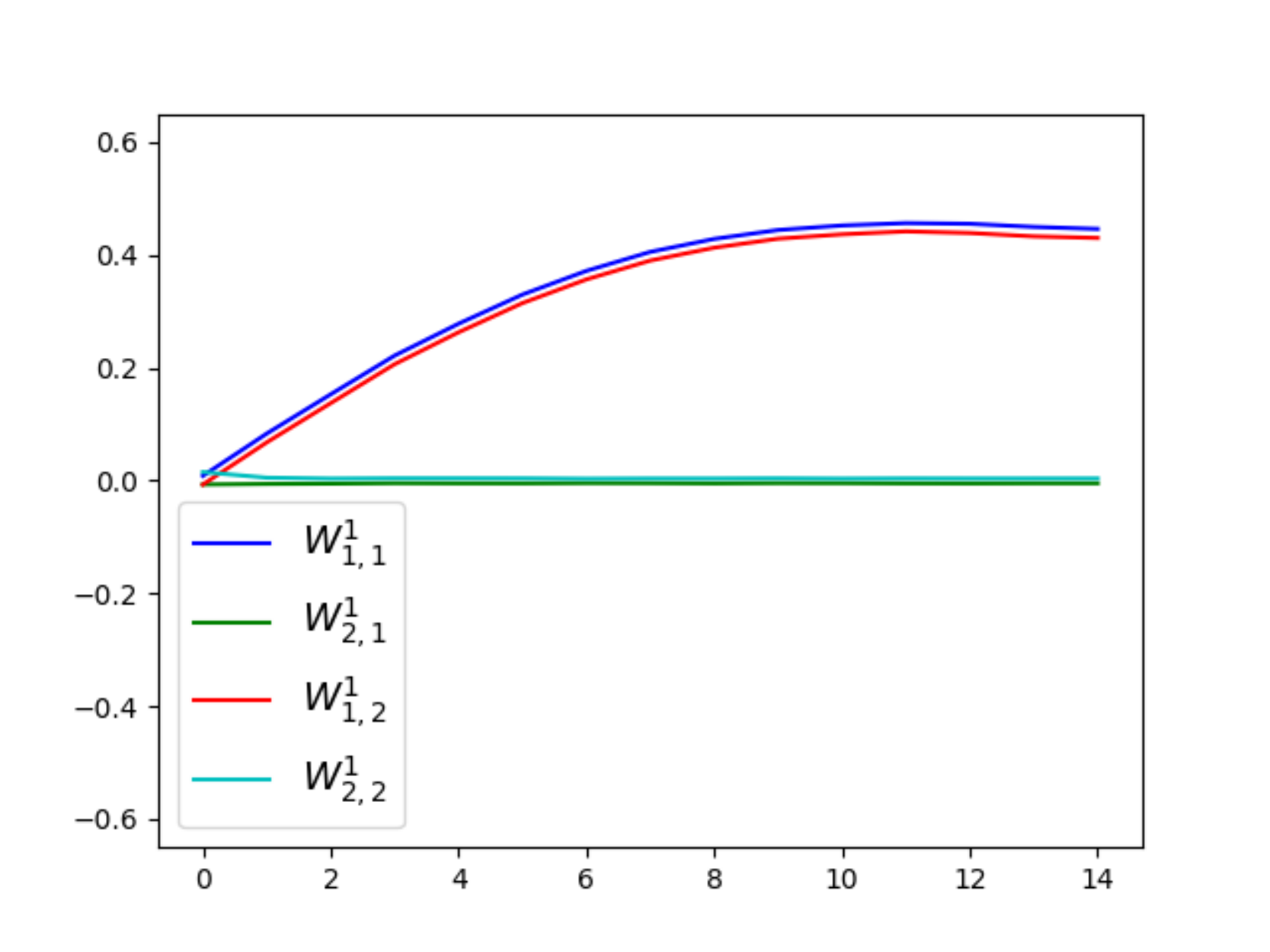} &
     \includegraphics[width=0.45\textwidth]{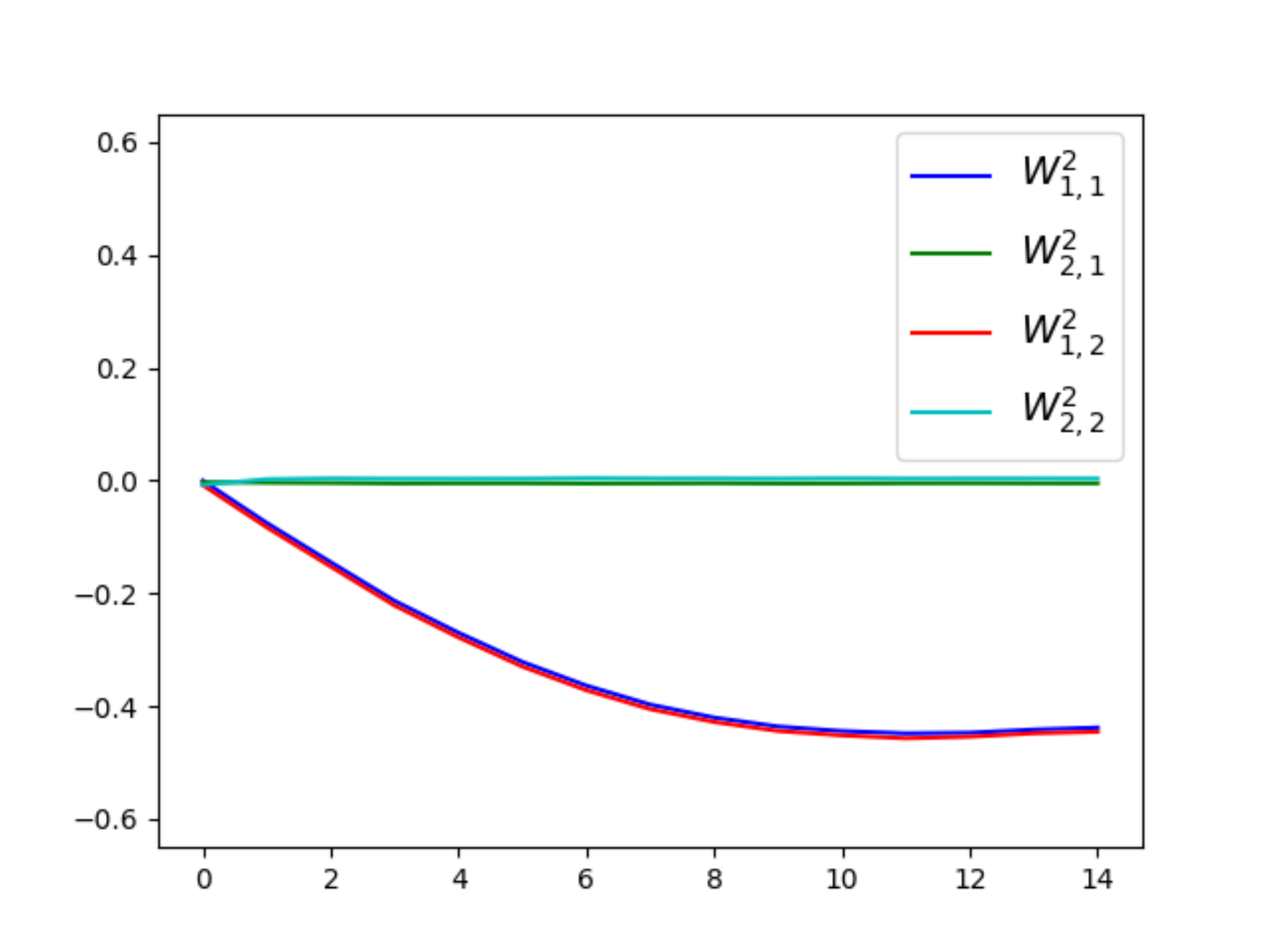} \\
     (a)    & (b) 
\end{tabular}
\caption{(a) the parameter values of $\mathbf{W}^1$ associated with $z_t$ for Agent 1  and (b) the parameter values of $\mathbf{W}^2$ associated with $z_t$ for Agent 2 (during the training phase)}
\label{fig:ToyExample_value_para}
\end{center}
\end{figure}

With this setup, we simply learned the policy parameters $\mathbf{W}^1$ and $\mathbf{W}^2$ associated with $z_t$ by greedily maximizing the instantaneous reward $r_t$ by stochastic gradient descent with Adam optimizer with learning rate $3 \times 10^{-4}$.  The parameter learning curves of $\mathbf{W}^1$ and $\mathbf{W}^2$ are shown in Fig. \ref{fig:ToyExample_value_para}.
It is observed that $w_{11}^1$ and $w_{12}^1$ of Agent 1 converge to positive values, whereas $w_{21}^1$ and $w_{22}^1$ of Agent 1 converge to zero. This setting of parameters $w_{11}^1$, $w_{12}^1$, $w_{21}^1$ and $w_{22}^1$ of Agent 1 generates movement of Agent 1 to the right side since $z_t^x \ge 0$ and $z_t^y \ge 0$ due to \eqref{eq:ztxDist} and \eqref{eq:ztyDist}. (Please see \eqref{eq:append_toy_a1}.)  On the other hand, $w_{11}^2$ and $w_{12}^2$ of Agent 2 converge to negative values, whereas $w_{21}^2$ and $w_{22}^2$ of Agent 2 converge to zero. This setting of parameters $w_{11}^2$, $w_{12}^2$, $w_{21}^2$ and $w_{22}^2$ of Agent 2 generates movement of Agent 2 to the left side since $z_t^x \ge 0$ and $z_t^y \ge 0$ due to \eqref{eq:ztxDist} and \eqref{eq:ztyDist}. (Please see \eqref{eq:append_toy_a2}.) Hence, the two agents meet. Note that the desired coordination between Agents 1 and 2 can be achieved by injecting common random variable $z_t$ and learning the set of parameters $\mathbf{W}^1$ and $\mathbf{W}^2$ associated with $z_t$ properly.


\begin{figure}[h]
\begin{center}
\begin{tabular}{cc}
      \includegraphics[width=0.45\textwidth]{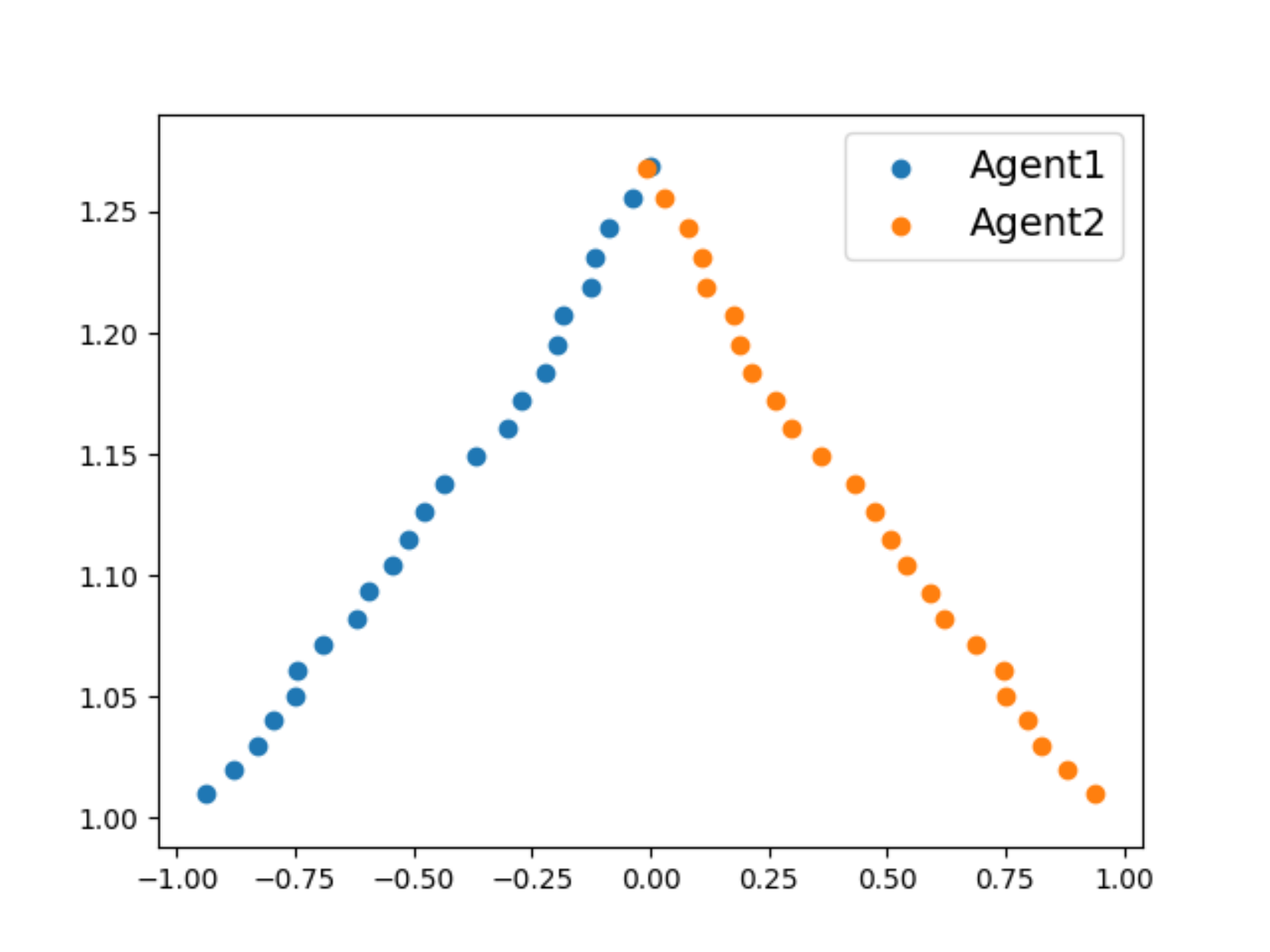} &
     \includegraphics[width=0.45\textwidth]{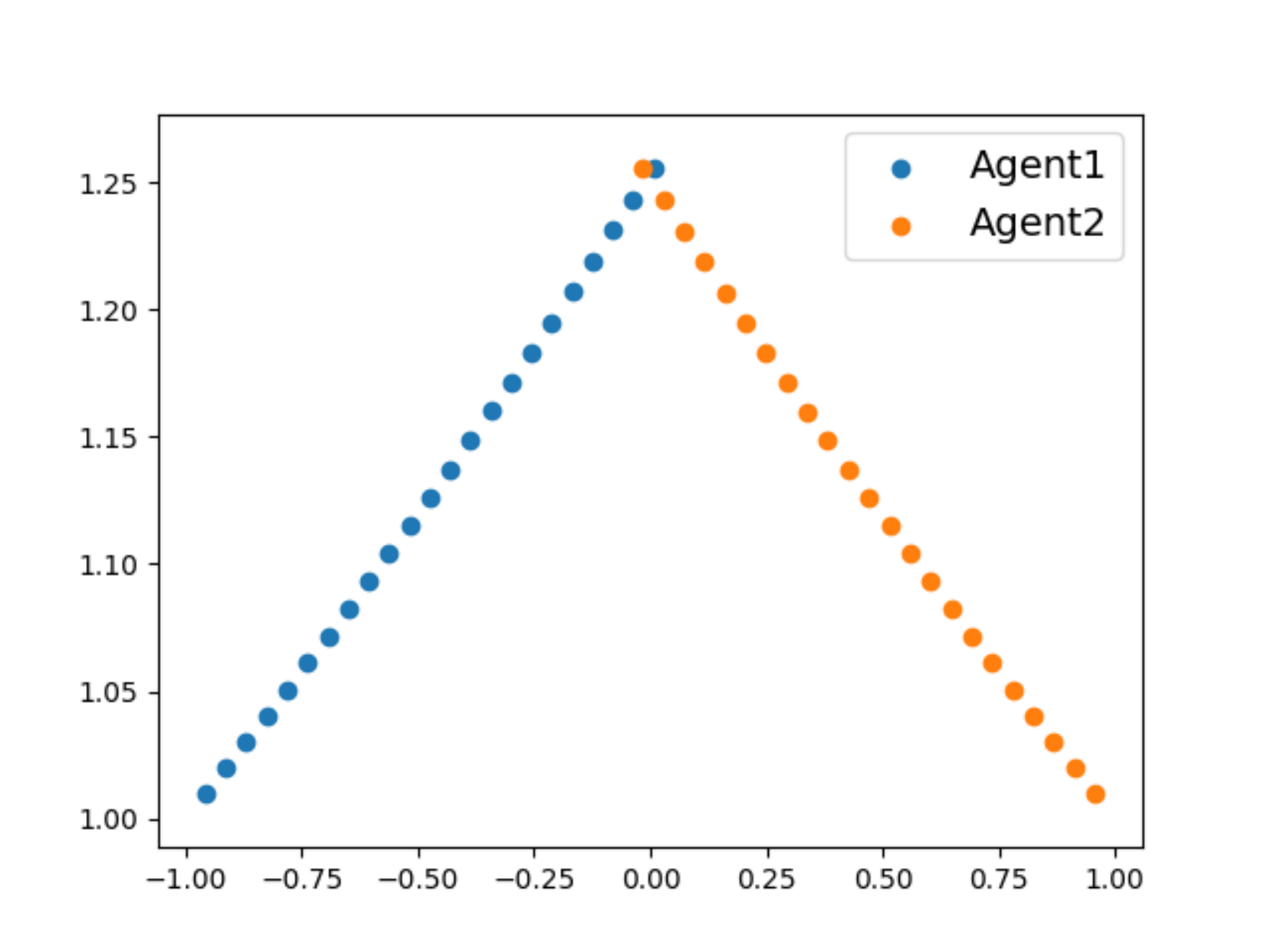} 
     \\
     (a)    & (b) 
\end{tabular}
\caption{The trajectory of two agents in the execution phase after training: (a) $z_t^x \sim \mbox{Unif}[0,0.1]$, $ z_t^y \sim \mbox{Unif}[0,0.1]$) and (b) $z_t^x = 0.05$, $z_t^y = 0.05$}
\label{fig:ToyExample_trajectory}
\end{center}
\end{figure}

Fig. \ref{fig:ToyExample_trajectory} shows the trajectories of Agents 1 and 2 for an episode in the execution phase after training.   Fig. \ref{fig:ToyExample_trajectory}(a) shows the trajectory when we input the random variable $z_t$ with distribution \eqref{eq:ztxDist} and \eqref{eq:ztyDist} to the policy network just as we did in the training phase.  Fig. \ref{fig:ToyExample_trajectory}(b) shows the trajectory when we input  $\mathbb{E}[z_t]=(0.05,0.05)$ to  the policy network for all $t$ in the execution phase after training. The desired action  is still obtained in the case of Fig. \ref{fig:ToyExample_trajectory}(b). This is because the parameters $\mathbf{W}^1$ and $\mathbf{W}^2$ associated with $z_t$ are properly learned during the training phase by enhanced exploration of the joint state-action space based on correlated exploration due to $\mathbf{W}^1z_t$ and $\mathbf{W}^2z_t$ with $z_t$ random.  This  learned parameters are used in the execution phase. We can view that by setting $z_t = \mathbb{E}[z_t]$, we pick and apply the representative joint bias on actions. Note that the desired joint bias in the case of Fig. \ref{fig:ToyExample_trajectory}(b) is obtained because of the fact that $z_t$ is distributed over $[0,0.1]$ by  \eqref{eq:ztxDist} and \eqref{eq:ztyDist}.  Hence, the choice of $P_Z$ is important in this method. However, at least the shift of the support of $z_t$ is not a big concern when a general neural network is used as the policy function. In the case of a general neural network as the policy function, shift of $z_t$ is automatically done by the node bias of the neural network and this node bias is also learned as parameter.

In this example, we observe that coordination of actions and coordinated exploration are feasible by injecting a common random variable $z_t$ to the input of every policy function and learning the parameters associated with $z_t$. In this example, we fixed the weights associated with the observation $o_t^i$ to show the exploration and control capability of the $z_t$ part. In  general cases, we have the freedom to design the weights associated with the observation $o_t^i$ too. Designing  the conventional policy 
parameters associated with the observation 
together with additional degree-of-freedom for exploration and design generated by injecting $z_t$  combined with nonlinear deep neural network can lead to learning of complicated coordinated behavior via correlated exploration. This paper fully develops this idea.

\newpage
\section*{Appendix B: Proofs}
\label{sec:appendProof}

In the main paper, we defined the state and state-action value functions for Agent $i$ as follows:
\begin{align}
    &Q_i^{\boldsymbol{\pi}}(s,a) \triangleq \mathbb{E}_{\scriptsize\begin{array}{c}
       \tau_0 \sim \boldsymbol{\pi}\\
        z_t\sim p_Z
        \end{array}}\Bigg[ r_0 + \gamma V_i^{\boldsymbol{\pi}}(s_{1}) \Bigg| s_0=s, \boldsymbol{a}_0=\boldsymbol{a} \Bigg], \label{eq:appendix_value_function11}\\
    &V_i^{\boldsymbol{\pi}}(s)
    \triangleq \mathbb{E}_{\scriptsize\begin{array}{c}
       \tau_0 \sim \boldsymbol{\pi}\\
        z_t\sim p_Z
        \end{array}}\Bigg[ \sum_{t=0}^{\infty}\gamma^t \Big(r_{t} +\beta H(a_t^i| s_{t}) \nonumber  +\frac{\beta}{N}\sum_{j\neq i}\log q^{(i,j)}(a_{t}^i,a_{t}^j|s_{t})\Big) \Bigg| s_0=s \Bigg], \label{eq:appendix_value_function12} \\
\end{align}
Then, the Bellman operator corresponding to  $V_i^{\boldsymbol{\pi}}$ and $Q_i^{\boldsymbol{\pi}}$ on the value function estimates $V_i(s)$ and $Q_i(s,\boldsymbol{a})$ is given by
\begin{align}\label{eq:bellmanAppend}
    \mathcal{T}^{\boldsymbol{\pi}}Q_i(s,\boldsymbol{a}) &\triangleq r(s,\boldsymbol{a}) + \gamma \mathbb{E}_{s'\sim p}[V_i(s')],
\end{align}
where
\begin{align}\label{eq:bellman2Append}
    &V_i(s)= \mathbb{E}_{\scriptsize\begin{array}{c}  \boldsymbol{a}\sim \boldsymbol{\pi}\\
        z_t\sim p_Z
        \end{array}}\Bigg[Q_i(s,\boldsymbol{a})
    -\beta \log \tilde{\pi}^i(a^i|s)+ \frac{\beta}{N} \sum_{j\neq i}\log q^{(i,j)}(a^i,a^j|s) \Bigg].
\end{align}

\eqref{eq:appendix_value_function11}, \eqref{eq:appendix_value_function11},
\eqref{eq:bellmanAppend} and
\eqref{eq:bellman2Append} are the rewritings of equations 
Equations (12), (13), (14) and (15) in the main paper.

\textbf{Proposition 1} 
(Variational Policy Evaluation). For fixed $\boldsymbol{\pi}$ and the variational distribution $q$, consider the modified Bellman operator $\mathcal{T}^{\boldsymbol{\pi}}$ in (\ref{eq:bellmanAppend}) and an arbitrary initial function $Q_i^{(0)}:\mathcal{S}\times\mathcal{A}\rightarrow \mathbb{R}$, and define $Q_i^{(k+1)}=\mathcal{T}^{\boldsymbol{\pi}}Q_i^{(k)}$. Then, $Q_i^{(k)}$ converges to $Q_i^{\boldsymbol{\pi}}$ defined in (\ref{eq:appendix_value_function11}).

\begin{proof}
	From \eqref{eq:bellmanAppend}, we have 
	\begin{align}
    	\mathcal{T}^{\pi}Q_i(s_t,\boldsymbol{a_t})&= r(s_t,\boldsymbol{a_t})     +\gamma \mathbb{E}_{\scriptsize
    	\begin{array}{c}
    	s_{t+1}\sim p,\boldsymbol{a_{t+1}}\sim \boldsymbol{\pi}\\\
    	z_{t+1}\sim p_Z
    	\end{array}} \Bigg[Q_i (s_{t+1},\boldsymbol{a_{t+1}})-\beta \log \tilde{\pi}^i(a_t^i|s_t) \nonumber \\
    	& ~~~~~~~~~~~~~~~~~~~~~~~~~~~~~~~~~~~~~~~~~~~~~~~~~~~~~~~~~~~~~~~~~~~~~~~~~ + \frac{\beta}{N} \sum_{j\neq i}\log q^{(i,j)}(a_t^i,a_t^j|s_t) \Bigg] \\
	    &= \underbrace{ r(s_t,\boldsymbol{a_t})     +\gamma  
	    \mathbb{E}_{\scriptsize
    	\begin{array}{c}
    	s_{t+1}\sim p,\boldsymbol{a_{t+1}}\sim \boldsymbol{\pi}\\\
    	z_{t+1}\sim p_Z
    	\end{array}}
	    \Bigg[-\beta \log \tilde{\pi}^i(a_t^i|s_t) + \frac{\beta}{N} \sum_{j\neq i}\log q^{(i,j)}(a_t^i,a_t^j|s_t)\Bigg]}_{=:r_{\pi}(s_t,\boldsymbol{a_t})} \nonumber \\
	    &~~~~~~~~+\gamma  \mathbb{E}_{\scriptsize
    	\begin{array}{c}
    	s_{t+1}\sim p,\boldsymbol{a_{t+1}}\sim \boldsymbol{\pi}\\\
    	z_{t+1}\sim p_Z
    	\end{array}} \Bigg[Q_i(s_{t+1},\boldsymbol{a_{t+1}})\Bigg]  \\
	    &= r_{\pi}(s_t,\boldsymbol{a_t})+\gamma  \mathbb{E}_{\scriptsize
    	\begin{array}{c}
    	s_{t+1}\sim p(\cdot|s_t,a_t),z_{t+1}\sim p_Z, \boldsymbol{a_{t+1}}\sim \boldsymbol{\pi}(\cdot|s_{t+1},z_{t+1})\\
    	\end{array}}\Bigg[Q_i(s_{t+1},\boldsymbol{a_{t+1}})\Bigg],
	\end{align}
	
	where in the last line the expectation arguments are explicitly shown without abbreviation for clarity. 
	Then, we can apply the standard convergence results for policy evaluation. Define
	\begin{align}
    	\mathcal{T}^{\pi}(v) = \mathcal{R}^{\pi} + \gamma \mathcal{P}^{\pi}v
	\end{align}
	for $v = [Q(s,\boldsymbol{a})]_{s \in \mathcal{S}, \boldsymbol{a} \in \mathcal{A}}$.
	Then,  the operator $\mathcal{T}^{\pi}$ is a $\gamma$-contraction.
	\begin{align}
    	\|\mathcal{T}^{\pi}(v) - \mathcal{T}^{\pi}(u)\|_{\infty}
    	&= \| (\mathcal{R}^{\pi} + \gamma \mathcal{P}^{\pi}v) - (\mathcal{R}^{\pi} + \gamma \mathcal{P}^{\pi}u) \|_{\infty} \\
	    &= \| \gamma\mathcal{P}^{\pi}(v-u) \|_{\infty} \\
    	&\leq \|  \gamma\mathcal{P}^{\pi}\|_{\infty}   \| v-u\|_{\infty} \\
    	&\leq \gamma \| u-v \|_{\infty}
	\end{align}
	since $||\mathcal{P}^{\boldsymbol \pi}||_\infty \le 1$
Therefore,  the operator $\mathcal{T}^{\pi}$ has a unique fixed point by the contraction mapping theorem. Let   $Q_i^\pi(s,\boldsymbol{a})$ be this fixed point.  Since
	\begin{equation}
		\| Q_i^{(k)}(s,\boldsymbol{a}) - Q_i^\pi(s,\boldsymbol{a}) \|_{\infty} \leq \gamma \| Q_i^{(k-1)}(s,\boldsymbol{a}) - Q_i^\pi(s,\boldsymbol{a})  \|_{\infty} \leq \cdots \leq \gamma^{k} \| Q_i^{(0)}(s,\boldsymbol{a}) - Q_i^\pi(s,\boldsymbol{a}) \|_{\infty},
	\end{equation}
	we have
	\begin{align}
		\lim_{k \rightarrow \infty} \| Q_i^{(k)} (s,\boldsymbol{a}) - Q_i^\pi(s,\boldsymbol{a}) \|_\infty = 0
	\end{align}
	and this implies
	\begin{equation}
		\lim_{k \rightarrow \infty} Q_i^{(k)} (s,\boldsymbol{a}) = Q_i^\pi(s,\boldsymbol{a}), \quad \forall (s, \boldsymbol{a}) \in (\mathcal{S} \times \boldsymbol{\mathcal{A}}).
	\end{equation}
	
\end{proof}


We proved the variational policy evaluation in a finite state-action space. We can expand the result to the case of an infinite state-action space by assuming the followings:
\begin{itemize}
    \item Assume that Q functions for $\pi$ are in L infinity
    \item From \cite{1999vpe1}, L infinity is a Banach space
    \item From \cite{2018vpe2}, by the  Banach fixed point theorem, Q function should  converge to a unique point in L infinity space and that is the Q function of given $\pi$
\end{itemize}

\vspace{3ex}

\textbf{Proposition 2} (Variational Policy Improvement). Let $\pi_{new}^i$ and $q_{new}$ be the updated policy and the variational distribution from (\ref{eq:policy_improvement}). Then, $Q_i^{\pi^i_{new}, \pi^{-i}_{old}}(s,\boldsymbol{a})\geq Q_i^{\pi^i_{old}, \pi^{-i}_{old}}(s,\boldsymbol{a})$ for all $(s,\boldsymbol{a}) \in (\mathcal{S}\times \boldsymbol{\mathcal{A}})$. $(\pi^i_{k+1}, q_{k+1}) =$

\begin{align}
    & \mathop{\arg\max}_{\pi^i, q} \mathbb{E}_{\scriptsize \begin{array}{c}(a^i,a^{-i})\sim (\pi^i, \pi_{k}^{-i})\\
    z_k \sim p_Z\end{array}} \Bigg[Q_i^{\boldsymbol{\pi}_k}(s,\boldsymbol{a}) -\beta \log \tilde{\pi}^i(a^i|s)+  \frac{\beta}{N} \sum_{j\neq i}\log q^{(i,j)}(a^i,a^j|s)  \Bigg]. \label{eq:policy_improvement}
\end{align}

\begin{proof} Let us rewrite 
\eqref{eq:policy_improvement} to clarify that which terms are given and which terms are the optimization arguments. We use the subscript "old" for the given terms. Then,
  $\pi_{new}$ is updated as $(\pi^i_{new}, q_{new}) =$
	\begin{align}
    	\mathop{\arg\max}_{\pi^i,q} \mathbb{E}_{\scriptsize \begin{array}{c}(a^i,a^{-i})\sim (\pi^i, \pi_{old}^{-i})\\
    z_k \sim p_Z\end{array}} \Bigg[Q_i^{\boldsymbol{\pi}_{old}}(s_t,\boldsymbol{a}_t) -\beta \log \tilde{\pi}^i(a_t^i|s_t) + \frac{\beta}{N} \sum_{j\neq i}\log q^{(i,j)}(a_t^i,a_t^j|s_t) ) \Bigg].
	\end{align}
	Then, the following inequality is hold
	\begin{align}
    	&\mathbb{E}_{\scriptsize
    	\begin{array}{c}(a_t^i,a_t^{-i})\sim (\pi_{new}^i, \pi_{old}^{-i})\\
    	z_k \sim P_Z\end{array}} \Bigg[Q_i^{\boldsymbol{\pi}_{old}}(s_t,\boldsymbol{a}_t) -\beta \log \tilde{\pi}_{new}^i(a_t^i|s_t) + \frac{\beta}{N} \sum_{j\neq i}\log q^{(i,j)}_{new}(a_t^i,a_t^j|s_t) ) \Bigg] \\
    	&\geq \mathbb{E}_{\scriptsize\begin{array}{c}(a_t^i,a_t^{-i})\sim (\pi^i_{old}, \pi_{old}^{-i})\\z_k \sim p_Z\end{array}} \Bigg[Q_i^{\boldsymbol{\pi}_{old}}(s_t,\boldsymbol{a}_t) -\beta \log \tilde{\pi}_{old}^i(a_t^i|s_t) + \frac{\beta}{N} \sum_{j\neq i}\log q^{(i,j)}_{old}(a_t^i,a_t^j|s_t) ) \Bigg] \\
    	&= V^{\boldsymbol{\pi}_{old}}_i(s_t).
	\end{align}
From the definition of the Bellman operator,
	\begin{align}
    	Q_i^{\boldsymbol{\pi}_{old}}(s_t,\boldsymbol{a_t})
    	&= r(s_t,\boldsymbol{a_t}) + \gamma \mathbb{E}_{s_{t+1}\sim p}[V_i^{\boldsymbol{\pi}_{old}}(s_{t+1})] \\
	    &\leq r(s_t,\boldsymbol{a_t}) + \gamma \mathbb{E}_{s_{t+1}\sim p}\mathbb{E}_{\scriptsize\begin{array}{c}(a_{t+1}^i,a_{t+1}^{-i})\sim (\pi_{new}^i, \pi_{old}^{-i})\\
	    z_{t+1}\sim p_Z\end{array}} \Bigg[Q_i^{\boldsymbol{\pi}_{old}}(s_{t+1},\boldsymbol{a}_{t+1})  \nonumber \\
	    &  ~~~~~ -\beta \log \tilde{\pi}_{new}^i(a_{t+1}^i|s_{t+1}) + \frac{\beta}{N} \sum_{j\neq i}\log q_{new}^{(i,j)}(a_{t+1}^i,a_{t+1}^j|s_{t+1})  \Bigg] \\
	    &\leq r(s_t,\boldsymbol{a_t}) + \gamma \mathbb{E}_{s_{t+1}\sim p}\mathbb{E}_{\scriptsize \begin{array}{c}(a_{t+1}^i,a_{t+1}^{-i})\sim (\pi_{new}^i, \pi_{old}^{-i})\\
	    z_{t+1}\sim p_Z \end{array}} \Bigg[r(s_{t+1},\boldsymbol{a_{t+1}})  \nonumber \\
	    & ~~~~~ -\beta \log \tilde{\pi}_{new}^i(a_{t+1}^i|s_{t+1}) + \frac{\beta}{N} \sum_{j\neq i}\log q_{new}^{(i,j)}(a_{t+1}^i,a_{t+1}^j|s_{t+1}) \nonumber \\
	    & ~~~~~~~~~~~~~~~~~~~~~~~~~~~~~~~~~~~~~~~~~~~~~~~~~~~~~ + \gamma \mathbb{E}_{s_{t+2}\sim p}\left[ V_i^{\boldsymbol{\pi}_{old}}(s_{t+2})\right]\Bigg] \\
	    &\ \ \ \vdots \nonumber \\
    	&\leq Q_i^{\pi^i_{new}, \pi^{-i}_{old}}(s_t,a_t).
	\end{align}
\end{proof}

\newpage

\section*{Appendix C: Details of Centralized Training}
\label{sec:appendCentralizedTraining}

The value functions $V^i_{\psi_i}(\boldsymbol{x})$, $Q^i_{\theta_i}(\boldsymbol{x},\boldsymbol{a})$ are updated based on the modified Bellman operator defined in (13) and (14). The state-value function $V^i_{\psi_i}(\boldsymbol{x})$ is trained to minimize the following loss function:
\begin{equation}\label{eq:practical_value}
    \mathcal{L}_V(\psi^i)=\mathbb{E}_{s_t\sim D}\left[ \frac{1}{2}(V^i_{\psi^i}(\boldsymbol{x}_t)-\hat{V}^i_{\psi^i}(\boldsymbol{x}_t))^2 \right]
\end{equation}
where $D$ is the replay buffer that stores the transitions $(\boldsymbol{x}_t,\boldsymbol{a}_t, r_t, \boldsymbol{x}_{t+1})$;  $Q^i_{min}(\boldsymbol{x}_t,a_t^i) = \text{min}[Q^i_{\theta^{i,1}}(\boldsymbol{x}_t,a_t^i), Q^i_{\theta^{i,2}}(\boldsymbol{x}_t,a_t^i)]$ is the minimum of the two action-value functions to prevent the overestimation problem \cite{2018Fujimoto}; and 
\begin{align} 
\hat{V}^i_{\psi^i}(\boldsymbol{x}_t) = \mathbb{E}_{z_t\sim N(0,\boldsymbol{I}),\{a^k\sim \pi^k(\cdot|o^k_t,z_t)\}_{k=1}^{N}}\Bigg[& Q^i_{min}(\boldsymbol{x}_t,\boldsymbol{a}_t)-\beta \log \pi^i_{\phi^i}(a_t^i|o^i_t,z_t) \nonumber \\ & + \frac{\beta}{N} \sum_{j\neq i}\log q_{\xi^i}^{(i,j)}(a_t^i,a_t^j|o_t^i,o_t^j) \Bigg].
\label{eq:hatViAppend}
\end{align}
Note that in the second term of the RHS of \eqref{eq:hatViAppend}, originally we should have used the marginalized version, 

$-\beta \log \tilde{\pi}_{\phi^i}^i(a_t^i|o_t^i)=-\beta \log \mathbb{E}_{z_t\sim N(0,\boldsymbol{I})} [\pi_{\phi^i}^i(a_t^i|o_t^i,z_t)]$.

However, for simplicity of computation, we took the expectation  $\mathbb{E}_{z_t\sim N(0,\boldsymbol{I})}$ outside the logarithm. Hence, there exists Jensen's inequality-type approximation error. We observe that this approximation works well.

 The two action-value functions are updated by minimizing the loss
\begin{equation}\label{eq:loss_critic1}
    \mathcal{L}_Q(\theta^i)=\mathbb{E}_{(\boldsymbol{x}_t,\boldsymbol{a}_t)\sim D}\left[ \frac{1}{2}(Q_{\theta^i}(\boldsymbol{x}_t,\boldsymbol{a}_t)-\hat{Q}(\boldsymbol{x}_t,\boldsymbol{a}_t))^2 \right]
\end{equation}
where
\begin{equation}\label{eq:loss_critic2}
    \hat{Q}(\boldsymbol{x}_t,\boldsymbol{a}_t) =r_t(x_t,\boldsymbol{a_t}) + \gamma \mathbb{E}_{\boldsymbol{x}_{t+1}}[V_{\overline{\psi}^i}\boldsymbol({x}_{t+1})]
\end{equation}
and $V_{\overline{\psi}^i}$ is the target value network, which is updated by the exponential moving average method. We implement the reparameterization trick to estimate the stochastic gradient of policy loss. Then, the action of agent $i$ is given by $a^i=f_{\phi^i}(s;\epsilon^i,z)$, where $\epsilon^i \sim \mathcal{N}(0,\boldsymbol{I})$ and $z\sim \mathcal{N}(0,\boldsymbol{I})$. The policy for Agent $i$ and the variational distribution are trained to minimize the following policy improvement loss,
\begin{align}
        \mathcal{L}_{\pi^i, q}(\phi^i, \xi)=\mathbb{E}_{{\scriptsize \begin{array}{c}
        s_t\sim D,\\
        \epsilon^i\sim \mathcal{N},\\ z\sim \mathcal{N}
        \end{array}}}\Bigg[ -Q^i_{\theta^{i,1}}&(\boldsymbol{x}_t,\boldsymbol{a}) + \beta \log \pi^i_{\phi^i}({a}^i|o^i_t, z) \nonumber \\ &- \frac{\beta}{N} \sum_{j\neq i}\log q_{\xi^i}^{(i,j)}(\pi^i_{\phi^i}({a}^i|o^i_t, z),\pi^j_{\phi^j}({a}^j|o^j_t, z)|o_t^i,o_t^j) \Bigg] \label{eq:loss_actor_va}
\end{align}
where $q_{\xi^i}^{(i,j)}(\pi^i_{\phi^i}({a}^i|o^i_t, z),\pi^j_{\phi^j}({a}^j|o^j_t, z)|o_t^i,o_t^j)$
\begin{equation} \label{eq:qxiiij}
 = \underbrace{q_{\xi^i}(\pi^i_{\phi^i}({a}^i|o^i_t, z)|\pi^j_{\phi^j}({a}^j|o^j_t, z)|o_t^i,o_t^j)}_{(a)} \underbrace{q_{\xi^i}(\pi^j_{\phi^j}({a}^j|o^j_t, z)|\pi^i_{\phi^i}({a}^i|o^i_t, z)|o_t^i,o_t^j)}_{(b)}.
\end{equation}
Again, for simplicity of computation, we took the expectation  $\mathbb{E}_{z_t\sim N(0,\boldsymbol{I})}$ outside the logarithm for the second term in the RHS in \eqref{eq:loss_actor_va}.
Since approximation of the variational distribution is not accurate in the early stage of training and the learning via the term (a) in  \eqref{eq:qxiiij} is more susceptible to approximation error, we propagate the gradient only through the term (b)  in  \eqref{eq:qxiiij} to  make learning stable.
Note that minimizing $-\log q_{\xi^i}(a^j|a^i,s_t)$ is equivalent to minimizing the mean-squared error between $a^j$ and $\mu_{\xi^i}(a^i,o^i,o^j)$ due to our Gaussian assumption on the variational distribution.

\newpage
\section*{Appendix D: Pseudo Code}

\begin{algorithm}[h]
   \caption{VM3-AC (L=1)}
   \label{alg:VM3-AC}
\begin{algorithmic}
   \STATE \textbf{Centralized training phase}
   \STATE Initialize parameter $\phi^i, \theta^i, \psi^i, \overline{\psi}^i,  \xi^i, ~\forall  i\in \{1,\cdots, N\}$
   \FOR{$episode=1,2,\cdots$}
   \STATE Initialize state $s_0$ and each agent observes $o_0^i$
   \FOR{$t<T$ and $s_t \neq$ terminal}
   \STATE Generate $z_t \sim \mathcal{N}(0,I)$ and select action $a_t^i\sim \pi^i(\cdot|o_t^i,z_t)$ for each agent $i$
   \STATE Execute $\boldsymbol{a_t}$ and each agent $i$ receives $r_t$ and $o_{t+1}^i$
   \STATE Store transitions in $D$
   \ENDFOR
   \FOR{each gradient step}
   \STATE Sample a minibatch from D and generate $z_l \sim \mathcal{N}(0,I)$ for each transition.
   \STATE Update $\theta^i, \psi^i$ by minimizing the loss (\ref{eq:loss_critic1}) and (\ref{eq:loss_critic2})
   \STATE Update $\phi^i, \xi^i$ by minimizing the loss (\ref{eq:loss_actor_va})
   \ENDFOR
   \STATE Update $\overline{\psi}^i$ using the moving average method
   \ENDFOR
   \STATE
   \STATE \textbf{Decentralized execution phase}
   \STATE Initialize state $s_0$ and each agent observes $o_0^i$
   \FOR {each environment step}
   \STATE Select action $a_t^i\sim \pi^i(\cdot|o_t^i,z_t)$ where $z_t=\overrightarrow{0}$ (or sample from the Gaussian random sequence generator with the same seed)
   \STATE Execute $\boldsymbol{a_t}$ and each agent $i$ receives $o_{t+1}^i$
   \ENDFOR
\\
\end{algorithmic}
\end{algorithm}

\newpage

\section*{Appendix E: Hyperparameter and Training Detail}

The hyperparameters for MA-AC,  MA-SAC, MADDPG, and VM3-AC are summarized in Table \ref{table:app1}.

\begin{table}[h]
\caption{Hyperparameters of all algorithms}
\label{table:app1}
\vskip 0.15in
\begin{center}
\begin{small}
\begin{sc}
\begin{tabular}{lcccccr}
\toprule
& MA-AC &  SI-MOA &MAVEN & MADDPG & VM3-AC \\
\midrule
Replay buffer size    & $5\times 10^5$ &  $5\times 10^5$ & $5\times 10^5$ & $5\times 10^5$  & $5\times 10^5$\\
Discount factor & 0.99   & 0.99 & 0.99 & 0.99 & 0.99\\
Mini-batch size  & 128  & 128  & 128 & 128 & 128\\
Optimizer       & Adam    & Adam & Adam& Adam& Adam \\
Learning rate & 0.0003  & 0.0003 & 0.0003 & 0.0003 & 0.0003 \\
Target smoothing coefficient & 0.005 & 0.005  & 0.005 & 0.005 & 0.005 \\
Number of hidden layers (all networks)  & 2 & 2 &  2 & 2 & 2 \\
Number of hidden units per layer & 128 & 128 & 128 & 128 & 128 \\
Activation function for hidden layer &  ReLU &  ReLU & ReLU & ReLU & ReLU \\
Activation function for final layer &  Tanh &  Tanh & Tanh & Tanh & Tanh \\
\bottomrule
\end{tabular}
\end{sc}
\end{small}
\end{center}
\vskip -0.1in
\end{table}

\begin{table}[h]
\caption{The temperature parameter $\beta$ and the dimension of the latent variable $z$ for VM3-AC on the considered environments. Note that the temperature parameter $\beta$ in I-SAC and MA-SAC controls the relative importance between the reward and the entropy, whereas the temperature parameter $\beta$ in VM3-AC controls the relative importance between the reward and the mutual information. }
\label{table:ablation}
\vskip 0.15in
\begin{center}
\begin{small}
\begin{sc}
\begin{tabular}{lcccr}
\toprule
VM3-AC & $\beta$ & Dim(z) \\
\midrule
MW (N=3)    & 0.05 & 8\\
MW (N=4)    & 0.1 & 8\\
PP (N=2)    & 0.15 & 8\\
PP (N=3)    & 0.1 & 8\\
PP (N=4)    & 0.2 & 8\\
CTC (N=4)    & 0.05 & 10\\
CTC (N=5)    & 0.05 & 10 \\
CN (N=3)    & 0.1 & 8 \\
\bottomrule
\end{tabular}
\end{sc}
\end{small}
\end{center}
\vskip -0.1in
\end{table}

\newpage

\section*{Appendix F: Environment Detail}
\begin{figure*}[t]
\begin{center}
\begin{tabular}{cccc}
     \includegraphics[width=0.24\textwidth]{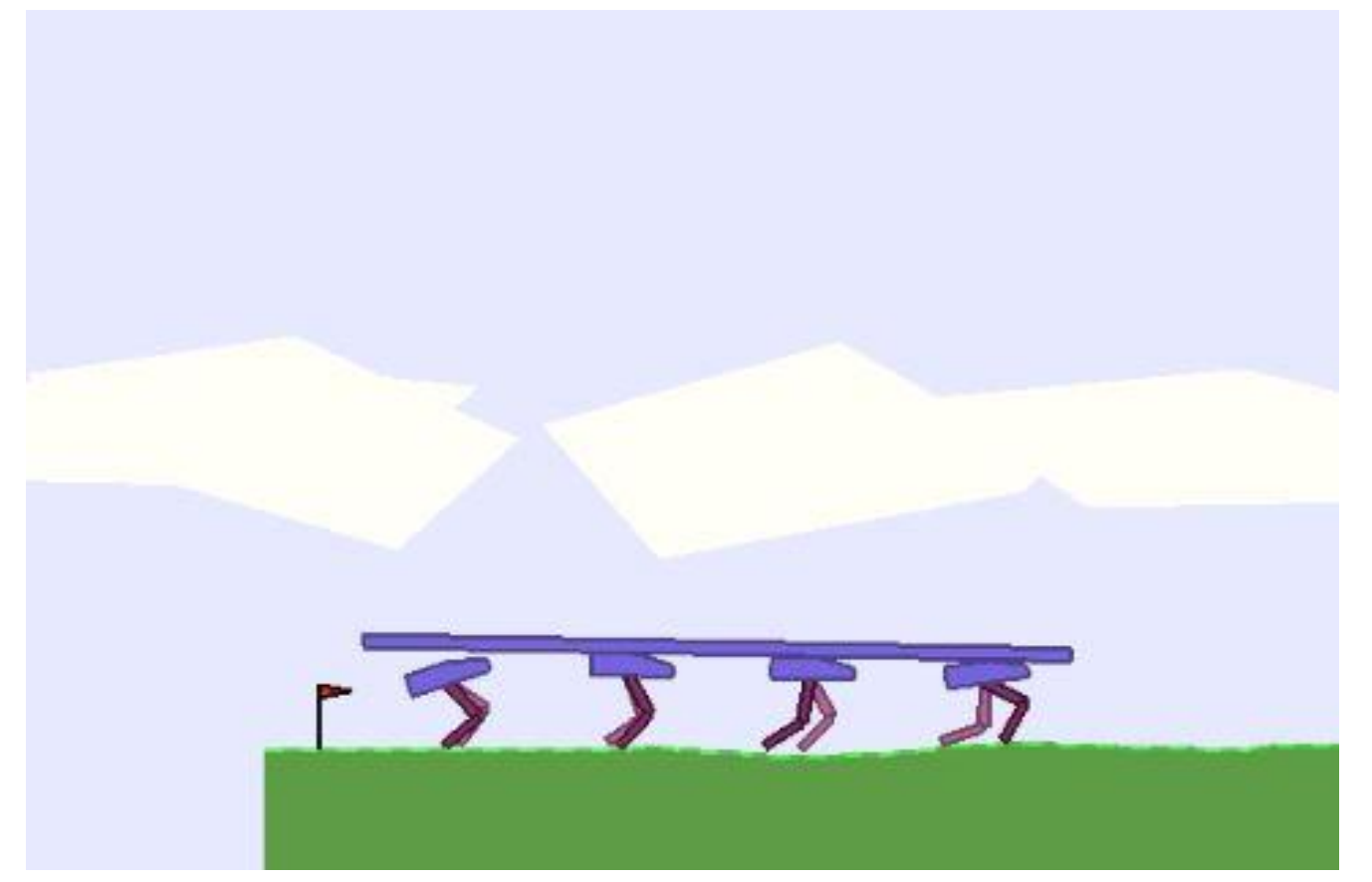} ~~~~~~~ &
     \includegraphics[width=0.15\textwidth]{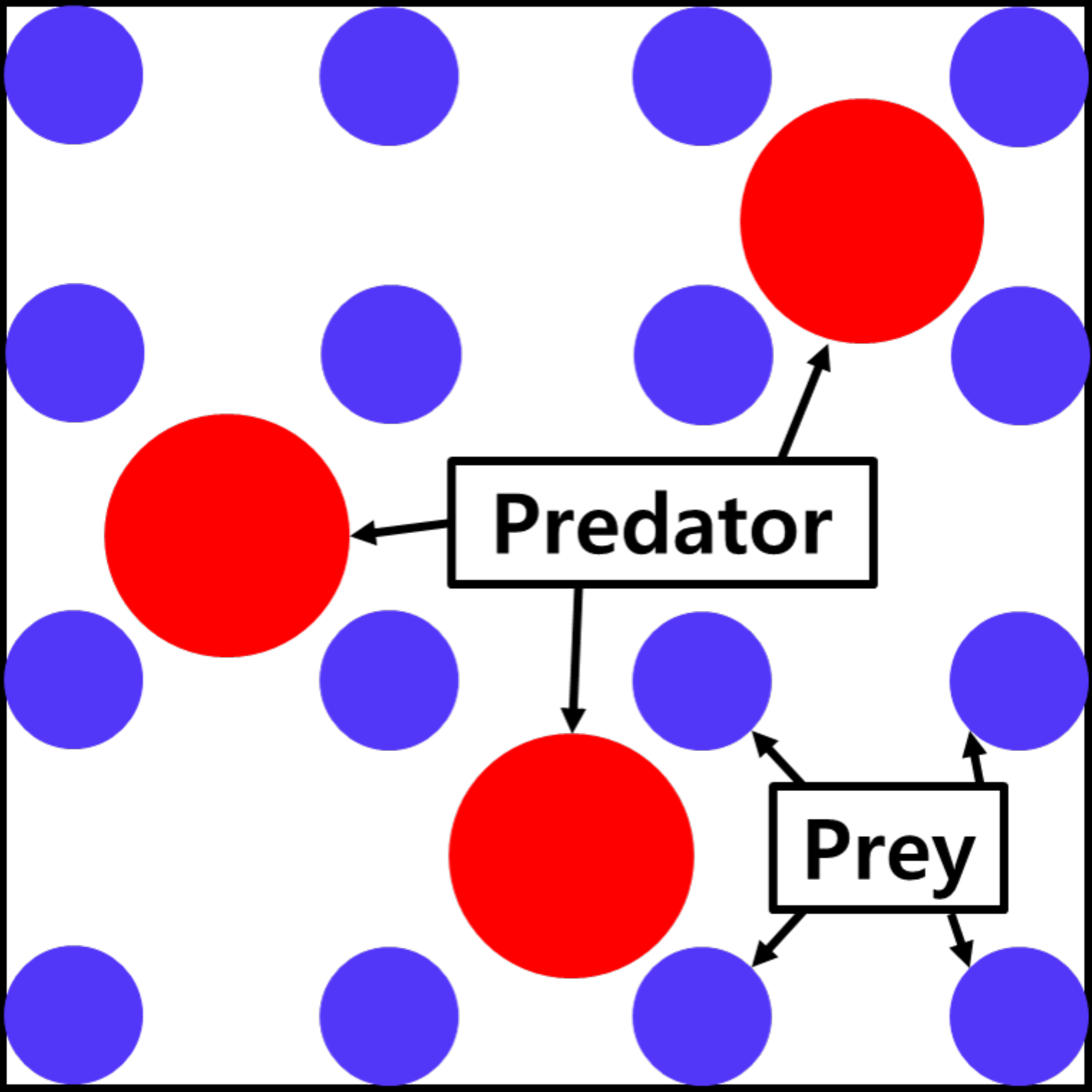} ~~~~~~~ &
     \includegraphics[width=0.15\textwidth]{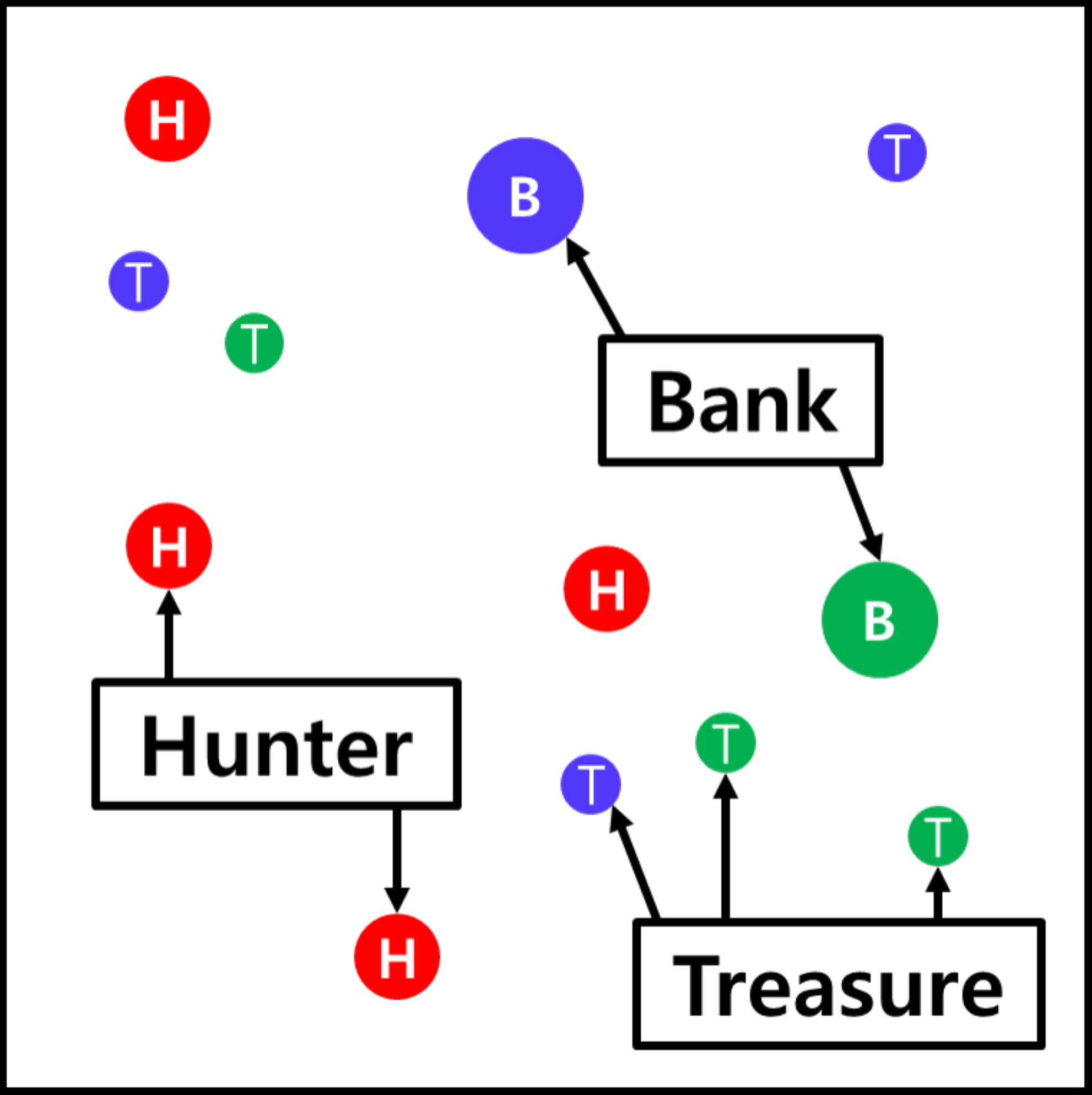} ~~~~~~~ &
     \includegraphics[width=0.15\textwidth]{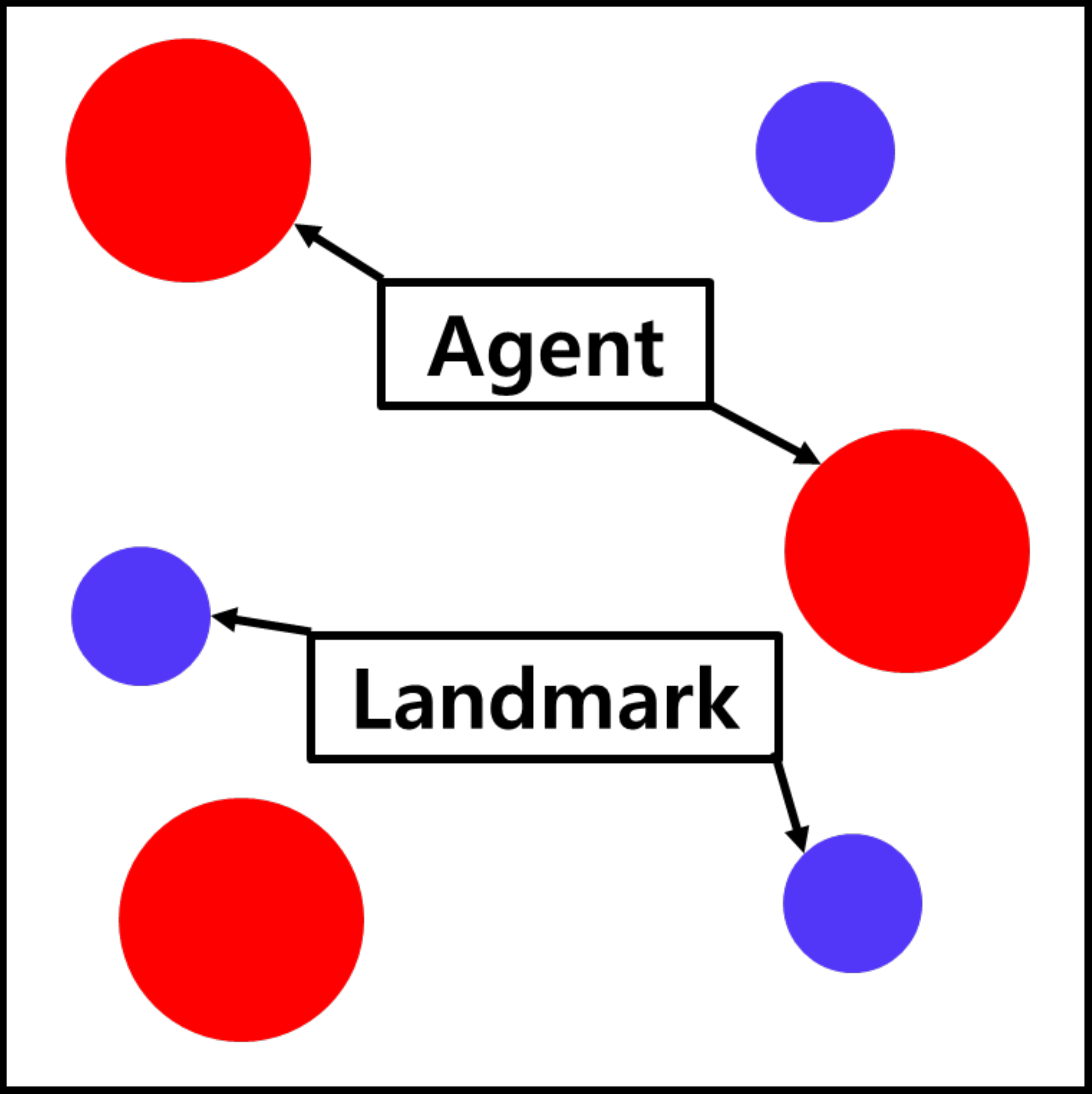} \\
     (a)~~~~~~~ & (b)~~~~~~~ & (c)~~~~~~~ & (d)
\end{tabular}
\caption{Considered environments: (a) Multi-Walker, (b) Predator-Prey, (c) Cooperative Treasure Collection, and (d) Cooperative Navigation}
\label{fig:environmentMain}
\end{center}
\end{figure*}

We implemented our algorithm based on OpenAI Spinning Up \cite{SpinningUp2018} and conduct the experiments on a server with Intel(R) Xeon(R) Gold 6240R CPU @ 2.40GHz. Each experiment took about 12 to 24 hours. We illustrate the considered environments in Fig. \ref{fig:environmentMain}.

\textbf{Multi-walker} The multi-walker environment, which was introduced in \cite{2018Gupta}, is a modified version of the BipedalWalker environment in OpenAI gym  to multi-agent setting. The environment consists of $N$ bipedal walkers and a large package. The goal of the environment is to move forward together while holding the large package on top of the walkers.
The observation of each agent consists of the joint angular speed, the position of joints.
Each agent has 4-dimensional continuous actions that control the torque of their legs.
Each agent receives  shared reward $R_1$ depending on the distance over which the package has moved and receives  negative local compensation $R_2$ if the agent drops the package or falls to the ground.
 An episode ends when one of the agents falls,  the package is dropped or $T$ time steps elapse.
  To obtain higher rewards, the agents should learn coordinated behavior.
  For example, if one agent only tries to learn to move forward, ignoring  other agents, then  other agents may fall.
  In addition, the different coordinated behavior is required as the number of agents changes.
    We set $T=500$, $R_2=-10$ and $R_1=10d$, where $d$ is the distance over which the package has moved.
    We simulated this environment in three cases by changing the number of agents ($N=2$, $N=3$, and $N=4$).

All algorithms used  neural networks to approximate the required functions.
 We used the neural network architecture proposed in \cite{2019Kim} to emphasize the agent's own observation and action for centralized critics. For Agent $i$, we used the shared neural network for the variational distribution $q_{\xi^i}(a_t^j|a_t^i,o_t^i, o_t^j )$ for $j\in \{1,\cdots,N\} \backslash \{i\}$, and the network takes the one-hot vector which indicates $j$ as input.

\textbf{Predator-prey} The predator-prey environment, which is a standard task for MARL, consists of $N$ predators and $M$ preys. We used a variant of the predator-prey environment into the continuous domain. The initial positions on the predators are randomly determined, and those of the preys are in the shape of a square lattice. The goal of the environment is to capture as many preys as possible during a given time $T$. A prey is captured when $C$ predators catch the prey simultaneously. The predators get team reward $R_1$ when they catch a prey. After all of the preys are captured and removed, we set the preys to respawn in the same position and increase the value of $R_1$.
Thus, the different coordinated behavior is needed as $N$ and $C$ change. The observation of each agent consists of relative positions between agents and other agents and those between agents and the preys. Thus, each agent can access to all information of the environment state. 
The action of each agent is two-dimensional physical action. We set $R_1=10$ and $T=100$. We simulated the environment with three cases: $(N=2, M=16, C=1$), $(N=3, M=16, C=1)$ and $(N=4, M=16, C=2)$.

\textbf{Cooperative treasure collection} The cooperative treasure collection environment, which was introduced in \cite{maac}, consists of $2$ banks, $N-2$ collectors, and $6$ hunters. Each bank has a different color and each treasure has one of the banks' colors. The goal of this environment is to deposit the treasures by controlling the banks and hunters. The hunters collect the treasure and then give it to the corresponding bank. Both hunters and banks receive shared reward $R_1$ if a treasure is deposited. The hunters receive a positive reward $R_2$ when a treasure is collected and a negative reward $-R_3$ if colliding with other agents. The observation of each agent consists of the locations of all other agents and landmarks, and action is two-dimensional physical action. We set $R_1=5$, $R_2=5$, $R_3=5$. We simulated the environment with two cases: $(N=4$) and $(N=5)$.

\textbf{Cooperative navigation} Cooperative navigation, which was proposed in \cite{2017Lowe}, consists of $N$ agents and $L$ landmarks. The goal of this environment is to occupy all landmarks while avoiding collision with other agents. The agent receives shared reward $R_1$ which is the sum of the minimum distance of the landmarks from any agents, and the agents who collide each other receive negative reward $-R_2$. In addition, all agents receive $R_3$ if all landmarks are occupied. The observation of each agent consists of the locations of all other agents and landmarks, and action is two-dimensional physical action. We set $R_2=10$, $R_3=1$, and $T=50$. We simulated the environment in the cases of ($N=3$, $L=3$). 

\newpage

\section*{Appendix G: SMAC environment}

We modified the SMAC environment to be sparse to make the problem more difficult. The considered sparse reward setting consists of a time-penalty reward which is obtained $-0.1$ every time step and a dead reward which is obtained $+10$ and $-1$ when one enemy dies and one ally dies, respectively. If all enemies die, the dead reward is given $+200$.

We implemented VM3-AC by modifying the code provided by \cite{FOP}. We replace the entropy term in \cite{FOP} with the sum of entropy and variational approximation. We used the categorical distribution with the dimension of $3$ for the latent variable. We used the deep neural network which consists of a 64-dimensional MLP with ReLU activation function, GRU, and an MLP to parameterize the policies. In addition, we use an MLP with 2 hidden layers which have 64 hidden units, and a ReLU activation function for both the critic networks. For the variational approximation, $q(a^j|a^i, s)$, we use the deep neural network which takes Agent $i$'s action and outputs Agent $j$'s action. The variational approximation is a feed-forward network whose weight is the output of a hyper-network which is a deep neural network taking the global state as input. The hyper-network is implemented similar to the mixing network in QMIX \cite{2018rashid}.

As in \cite{FOP}, we annealed the temperature parameter from $0.5$ to $0.05$ over $2\times 10^5$ steps. We provided source code in the supplementary material.

\section*{Appendix H: Broader Impact}

The research topic of this paper is multi-agent reinforcement learning (MARL). MARL is an important branch in the field of reinforcement learning. 
MARL models many practical control problems in the real world such as smart factories, coordinated robots, and connected self-driving cars. With the advance of knowledge and technologies in MARL, solutions to such real-world problems can be improved and more robust. For example, if the control of self-driving cars is coordinated among several nearby cars, the safety involved in self-driving cars will be improved much. So, we believe that the research advances in this field can benefit our safety and future society.  On the other hand, research advances in MARL, RL, and AI, in general, may affect the job situation. Some jobs may disappear and other jobs may newly appear. But, this situation always occurred when new technology was invented, e.g. automobiles.


\end{document}